\documentclass{svjour3}
\usepackage{graphicx}
\usepackage{float}
\usepackage[justification=centering]{subfig}
\usepackage{cite}
\usepackage{hyperref}
\hypersetup{hidelinks}
\usepackage{url}
\usepackage{listings}
\usepackage{xcolor}
\usepackage{tcolorbox}
\usepackage{caption}
\usepackage{booktabs}
\usepackage{multirow}
\usepackage{enumerate}
\usepackage{textcomp}
\usepackage{booktabs}
\usepackage[sort,compress]{natbib}
\usepackage{diagbox}
\usepackage{enumitem}
\usepackage[misc,geometry]{ifsym}
\usepackage{longtable,tabu}
\usepackage{afterpage}
\usepackage{verbatim}
\usepackage{xurl}
\usepackage{tikz} 
\usepackage{rotating}
\usepackage[toc,page]{appendix}
\usepackage{graphicx}
\usepackage[T1]{fontenc}
\usepackage{pdflscape}
\usepackage{caption} 
\usepackage{subfloat}
\usepackage{longtable}
\usepackage{ltxtable}
\usepackage{fancyvrb}
\usepackage{tabularx}
\usepackage{framed}
\usepackage{fontawesome}
\usepackage{array} % for better column control

\newtcolorbox{qoutebox}[3][]
{
  colframe = gray!30!white,
  colback  = #2!10,
  #1,
}

\begin{document}

\title{How Do Users Revise Architectural Related Questions on Stack Overflow: An Empirical Study}
\titlerunning{Revise Architectural Related Questions on Stack Overflow}

\author{Musengamana Jean de Dieu \and
        Peng Liang      \and
        Mojtaba Shahin  \and
        Arif Ali Khan   %\and
        % Chen Yang      \and
        % Zengyang Li    \and
}

\institute{Musengamana Jean de Dieu \and Peng Liang (\Letter) \at School of Computer Science, Wuhan University, Wuhan, China \\ 
            %Hubei Luojia Laboratory, Wuhan, China\\
            \email{mjados@outlook.com, liangp@whu.edu.cn}
            \and
            Mojtaba Shahin \at School of Computing Technologies, RMIT University, Melbourne, Australia \\ 
            \email{mojtaba.shahin@rmit.edu.au}
            \and
            Arif Ali Khan \at M3S Empirical Software Engineering Research Unit, University of Oulu, Oulu, Finland \\ 
            \email{arif.khan@oulu.fi}
            % \and
            % Chen Yang \at School of Artificial Intelligence, Shenzhen Polytechnic, Shenzhen, China \\ 
            % State Key Laboratory for Novel Software Technology, Nanjing University, Nanjing, China \\
            % \email{yangchen@szpt.edu.cn}
%             \and 
%             Zengyang Li \at School of Computer Science \& Hubei Provincial Key Laboratory of Artificial Intelligence and Smart Learning, Central
% China Normal University, Wuhan, China \\ 
%             \email{zengyangli@ccnu.edu.cn}
}

\date{Received: date / Accepted: date}

\maketitle

\begin{abstract}
Technical Questions and Answers (Q\&A) sites, such as Stack Overflow (SO), accumulate a significant variety of information related to software development in posts from users. To ensure the quality of this information, SO encourages its users to review posts through various mechanisms (e.g., question and answer revision processes). Although Architecture-Related Posts (ARPs) communicate architectural information that has a system-wide impact on development, little is known about how SO users revise information shared in ARPs. To fill this gap, we conducted an empirical study to understand how users revise Architecture-Related Questions (ARQs) on SO. By following a rigorous procedure consisting of data collection and filtering, we manually checked 13,205 ARPs and finally identified 4,114 ARQs that contain revision information. Our main findings are that: (1) The revision of ARQs is not prevalent in SO, and an ARQ revision starts soon after this question is posted (i.e., from 1 minute onward). Moreover, the revision of an ARQ occurs before and after this question receives its first answer/architecture solution, with most revisions beginning before the first architecture solution is posted. Both Question Creators (QCs) and non-QCs actively participate in ARQ revisions, with most revisions being made by QCs. (2) A variety of information (14 categories) is missed and further provided in ARQs after being posted, among which \textit{design context}, \textit{component dependency}, and \textit{architecture concern} are dominant information. (3) \textit{Clarify the understanding of architecture under design} and \textit{improve the readability of architecture problem} are the two major purposes of the further provided information in ARQs. (4) The further provided information in ARQs has several impacts on the quality of architecture solutions, including \textit{making architecture solution useful}, \textit{making architecture solution informative}, \textit{making architecture solution relevant}, among others. \textcolor{black}{The categorization of ARQ revisions was perceived by practitioners as meaningful, reliable, and comprehensive in capturing how ARQs evolve on SO. Participants confirmed that the categories effectively reflect common types, purposes, and impacts of ARQ revisions.} Our findings (1) provide researchers with new directions, such as the techniques and tools to facilitate SO users revising ARQs, and (2) provide SO users with suggestions and a structured template when asking and revising ARQs with the likelihood of getting quality architecture solutions.

\keywords{Architecture-Related Question \and Stack Overflow \and Question Revision  \and Architectural Information}

\end{abstract}

\section{Introduction} \label{introduction} 
Technical question and answer (Q\&A) sites, such as Stack Overflow (SO), are becoming an important and popular platforms for knowledge sharing and learning. SO has shown to be the most popular Q\&A website, and it is widely used by developers as well as architects to exchange software development related knowledge through asking and answering questions. One significant challenge for SO is to ensure the quality of the content on the site, given that users’ expertise, commitment, and experience are highly varied \citep{sheikhaei2023study}. %However, like at other Q\&A websites, asking and answering questions on SO may not always be straightforward even for experienced users \citep{zhou2020bounties}. For instance, questions on SO may lack the explanations for some important software development related concepts, which may in turn make such questions difficult to understand \citep{ford2018we} and consequently get no answers. 
For example, SO posts may contain incorrect explanations for some development related concepts or be full of spelling or grammar errors, which may make such posts difficult to read and comprehend. As a result, SO has developed several mechanisms to ensure the quality of the contents of its posts (e.g., revising or editing questions and answers). SO users revise their posts, for example, by adding, deleting, editing, or updating information in the contents of the posts (e.g., questions and answers). A major mechanism on SO to encourage users to revise the posts (i.e., questions and answers) is the use of a badge system \citep{anderson2013steering}. Users are awarded badges based on quantitative measures (e.g., by revising more than 500 questions or answers on SO)\footnote{\url{https://tinyurl.com/mwkmu7mu}}. 

Several empirical studies have extensively investigated how users revise programming related posts shared on Q\&A sites (such as SO) to ensure the quality of these posts and consequently enhance the development \citep{ford2018we, chen2017community, li2015good}. Yet, programming related pots from Q\&A sites represent just one facet of software development related posts that users share and revise on Q\&A sites. Users share and revise many other types of software development related posts on Q\&A sites. For example, users revise Architecture-Related Posts (ARPs) shared on Q\&A sites, such as SO. To the best of our knowledge, there has been no investigation of ARP revisions (e.g., ARQs) on Q\&A sites, such as SO, and it is unknown how users revise these posts. To bridge this gap, we conducted an empirical study to provide a comprehensive understanding of ARQ revisions on SO. By following a rigorous procedure consisting of data collection and filtering (see Section \ref{Methodology}), we manually checked 13,205 ARPs and finally identified 4,114 ARPs that contained revision information. A comprehensive quantitative and qualitative analysis was conducted to study the prevalence of ARQ revisions, the missing and further information provided in ARQs after being posted, the purposes of the further information provided in ARQs, and the impact of the further information provided in ARQs. In summary, the \textbf{key findings} of our study are as follows:

\begin{enumerate}[leftmargin=3ex]
\item An ARQ revision starts soon after the ARQ is posted (i.e., from 1 minute onward). In addition, the revision of an ARQ occurs \textit{before} and \textit{after} the ARQ receives its first answer/architecture solution, with most revisions beginning before the first architecture solution is posted. Both Question \textit{Creators (QCs) and non-QCs} \textit{actively participate} in ARQ revisions, with most revisions being made by QCs. 
\item SO users miss and further provide a variety (14 categories) of information in ARQs after being posted, wherein \textit{design context}, \textit{component dependency}, and \textit{architecture concern} are dominant information. 
\item SO users provide further information during the ARQ revisions to serve different purposes, in which \textit{clarify the understanding of architecture under design} and \textit{improve the readability of architecture problem} are the two major purposes of the further provided information in ARQs.
\item Providing further information in an ARQ has several impacts on the quality of the answer/architecture solution, such as \textit{making the architecture solution helpful}, \textit{making the architecture solution informative}, and \textit{making the architecture solution relevant}, among others.
\end{enumerate}

In the following of the paper: Section \ref{MotivationAndExample} presents the motivation and example, and Section \ref{background} introduces the background of this study. Section \ref{relatedwork} summarizes the related work. Section \ref{Methodology} describes the methodology used in this study. Section \ref{Results} presents the study results. Section \ref{Discussion} discusses the key findings of our research questions with their implications. Section \ref{ThreatValidity} clarifies the threats to validity. Section \ref{Conclusions} concludes this study with future work. 
\color{black}

%Researchers investigated SO posts (i.e., questions and answers) from different perspectives. For example, Wang \textit{et al}. analyzed how the badge system impacts answer revision on SO and found that the current system fails to consider the quality of revision. Chen \textit{et al}. proposed a deep learning approach to help users on SO fix grammar errors \citep{chen2017community}. Zhu \textit{et al}. examined the collaborative editing of posts (i.e., both answers and questions) on SO, and explored its benefits on content quality and potential negative effects on users' activity \citep{li2015good}. %The abovementioned studies extracted posts from SO and investigated the revisions in its posts (i.e., questions and answers) from different aspects. However, no prior study has specifically investigated the revision of information provided in architecture-related posts (ARPs), i.e., architecture-related questions (ARQs) in SO in terms of types of missing and further provided information in ARQs, purposes of the further provided information in ARQs, and the impact of the further provided information in ARQs on the answers/architecture solutions, which is the focus of this study

\section{Motivating Example}\label{MotivationAndExample}

On SO, questions (such as ARQs or programming related questions) serve as a starting point for curating crowdsourced knowledge \citep{zhu2022empirical}. ARQs arise during development when addressing specific architecture design concerns (e.g., quality attributes) and their trade-offs \citep{SA2012}. Unlike programming related questions, ARQs or problems communicate architectural information that has a system-wide impact on development \citep{SA2012}. Moreover, unlike asking and answering other types of questions (e.g., programming related questions), asking and answering ARQs on SO may not always be straightforward, even for experienced users. ARQs are often hard to frame or ask due to, for example, a vast amount of information (e.g., design contexts, architecture patterns) that need to be elaborated and clarified in the questions. %We observed some scenarios on SO where some users claimed that asking ARQs is not easy for them. 
SO users might claim that asking ARQs is not easy for them. One concrete example is this ARQ (SO post\#829597) in which a user (an asker) asked about an approach for designing an architecture of a system that will use plugins to achieve a higher level of reusability. Another user (a commenter) was not happy with the question and commented in the comment thread of this question: “\textit{At the risk of sounding glib, you need to try to make this question shorter}”. The asker replied in the comment thread: “\textit{(...) Sorry for the long post, it’s quite difficult to describe my problem}”. Furthermore, an ARQ on SO may be ambiguous or lack certain significant information (e.g., lacking detailed information about component dependencies), which may, in turn, make such questions difficult to understand and consequently get non-quality answers or no answers at all. For example, in this ARQ (SO post\#42440555), a user was not happy with the question description because the question failed to elaborate on module dependencies and provide an architectural diagram. A user (i.e., commenter) posted a comment asking to add more information in the question body: “\textit{Your question cannot be answered without making (many) assumptions. More information is needed about the dependencies of the modules. Are their stateless? Can you draw a flow (diagram) of the requests? (...)}”. Then an asker revised the question and replied to that comment: “\textit{Hi. Thanks for the comment. I have edited the description. Regards!}”. %For example, Figure \ref{ExampleOfQuestionRevisionThread} shows an ARP (SO post\#29644916) (including question body and comment thread) in which SO users discussed and revised this post. Specifically, an asker posted an ARQ asking about Microservice authentication strategies. This question was asked on April 15, 2015, at 8:10. However, later on the same date, just after almost four hours (e.g., April 15, 2015, at 12:29), another user (i.e., commenter) was not satisfied with the information provided in this question because the question was missing some important information (specifically, the asker failed to clarify the architecture concern and in use database system in the question body) by stating that: “\textit{Would you please provide more details on what you are trying to achieve? In the first case does authentication happen against Redis, or in the services themselves? Redis is missing in the second diagram (...)}”. Then the asker revised the question and stated in the comment: “\textit{I have added some information. Please let me know if it is still unclear. Thanks.}”

Similar to the revision of information shared in programming related posts, information shared in ARPs, such as ARQs, is revised too. However, as it stands, there is still a lack of understanding of the revision activities of ARQs on Q\&A sites, such as SO, in software development. Consequently, this hinders researchers from being aware of the potential challenges users (e.g., developers) face when asking and revising ARQs on Q\&A sites. It also creates difficulty for users to know the types of missing and further provided information in ARQs, and the purposes of the further provided information in ARQs after being posted. Moreover, it prevents users from knowing the impact of the further provided information in ARQs on the answers/architecture solutions. By understanding how users revise ARQs on Q\&A sites, such as SO, we can provide insights to researchers for improving the ways ARQs are asked and revised on SO (e.g., proposing approaches and tools to facilitate the revisions of ARQs on SO). In addition, it can also help the owners and users of SO improve the ways ARQs are asked and revised on SO, which would increase the likelihood of getting quality answers/architectural solutions on SO, and consequently improve knowledge sharing and speed up the development. Thus, this motivated us to carry out a study aiming to bridge this gap.

\section{Background} \label{background}
In this section, we give a brief overview of SO and the revision of ARQs on SO by using concrete examples from SO.

\subsection{Stack Overflow}
\begin{figure}
 \centering
\includegraphics[width=1\linewidth]{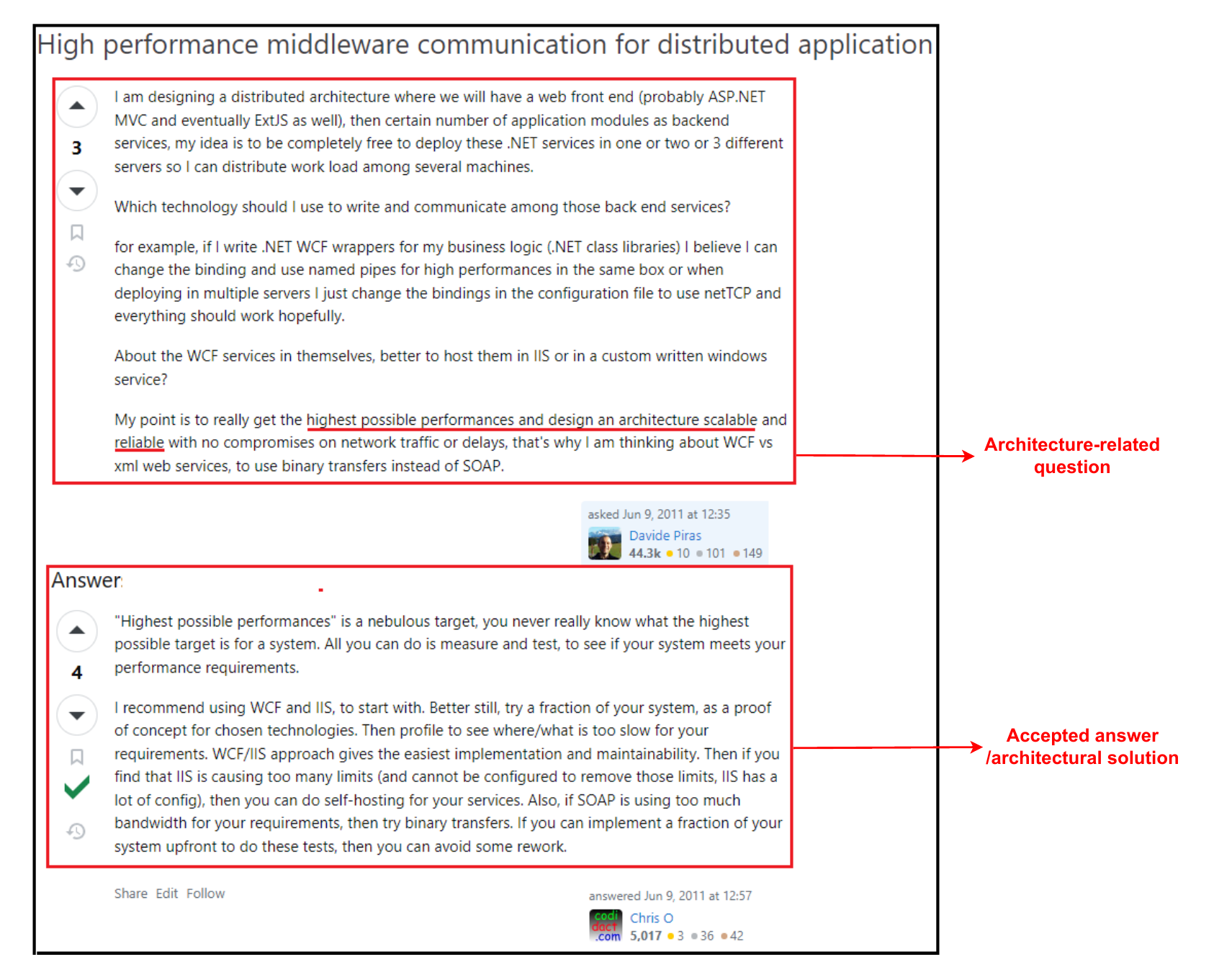}
 	\caption{An example of an ARQ and its accepted answer/architectural solution on Stack Overflow.} %The example highlights several details, such as a user who asked an ARQ (i.e., an asker), question creation date, question last edited date, question comment, answer creation date, and answer comments.
  \label{ExampleOfARQAnswerAndComments}
\end{figure}

SO is a technical Q\&A website where users (e.g., developers and architects) ask, answer, and discuss questions related to software development \citep{terragni2021apization, gao2023know, wu2023leveraging, de2023characterizing, soliman2018improving, soliman2017developing}. This site has been widely adopted by the software engineering community and has become the largest public information base for software development-related issues. Among other types of development information discussed in SO, architectural information, such as architectural patterns and architectural tactics \citep{bi2021mining, de2023characterizing}, has been shared in SO to assist architecture design. For example, Figure \ref{ExampleOfARQAnswerAndComments} shows an SO post \#6292867, in which a user requested technologies (including Windows Communication Foundation (WCF) and Simple Object Access Protocol (SOAP)) for designing a high performance, scalable, and reliable distributed architecture. The answerer of the post suggested and elaborated certain frameworks as solutions to the question, including WCF. %Figure \ref{ExampleOfARQAnswerAndComments} shows an SO post \#75826522, in which users discussed and shared architecture information including Model-View-ViewModel (MVVM) and Model–View–Presenter (MVP) architecture patterns. Specifically, a user (asker) asked to know if using MVVM and Clean architecture will improve performance of Android application and how such performance could be achieved (i.e., an architecture tactic to follow for performance improvement). The answer/architecture solution of the post stated and explained what (such as business logic code) should be optimized for performance improvement. 

\subsection{Revisions of ARQs on Stack Overflow} 
\begin{figure}[h]
 \centering
\includegraphics[width=1.0\linewidth]{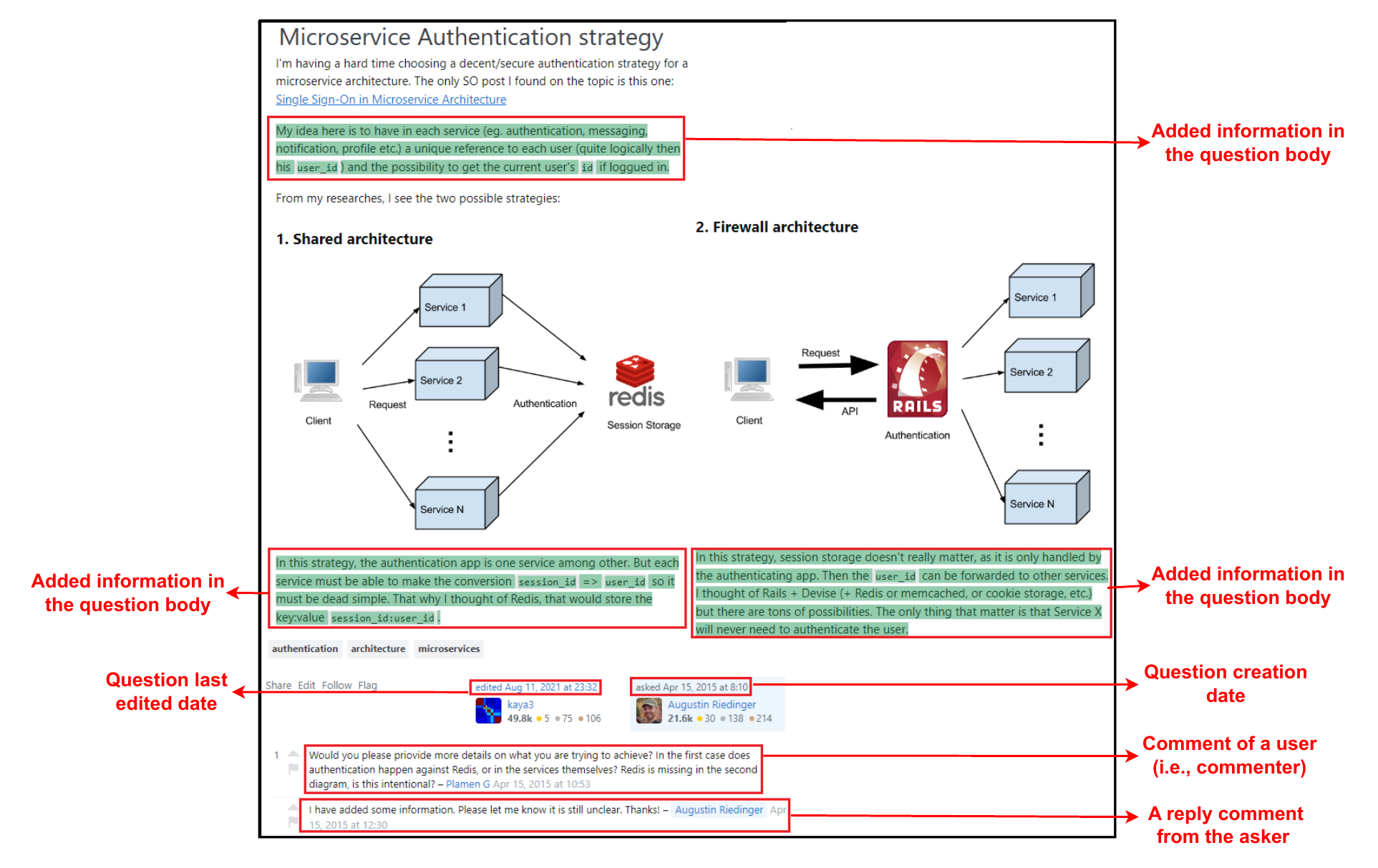}
 	\caption{An example of an ARQ revision (the added information is highlighted in green by Stack Overflow)}
  \label{ExampleOfQuestionRevisionThread}
\end{figure}

%Similar to other types of posts (e.g., programming related posts), 
SO allows users to post ARQs, answer those questions, and leave comments on either ARQs or answers/architectural solutions. In addition to the text descriptions, SO users may include architectural diagrams (e.g., component diagrams) or external sources (in the form of URL links) of architectural information in the contents of ARQs to enrich their questions. As mentioned earlier, asking ARQs in SO is not easy for some SO users (e.g., developers) due to various reasons. An ARQ in SO may be too broad, vague, or lack certain significant architectural information, which may in turn make such questions difficult to understand and consequently get non-quality answers or no answers at all. For example, this is an ARQ (SO post \#30559529) that was asked almost nine years ago (i.e., the question was posted on May 31, 2015 in SO) and the question did not receive any answer due to the reason that the question is too broad, as a user (a commenter) stated: “\textit{I think the question is perhaps too broad in scope for the design and implementation which is why it remains unanswered}”. Such unanswered ARQs can bring inconvenience to the askers (e.g., developers), which can not only hamper the efficiency of their work but also impoverish their development skills. In order to assist users (askers) ask their questions in an efficient way, including ARQs, SO allows askers or other users to revise ARQs after being posted. Specifically, after an asker posted an ARQ and this question was not well welcomed by other users due to various reasons, those users may request the asker to edit or clarify his/her question or those users may revise the question by themselves. %SO allows any registered user to perform revisions of the posts. Once a revision is performed by a user, the revision will be added to a review queue and wait for users with more than 2,000 reputation scores (reputation is a rough measurement of how much the community trusts a user\footnote{\url{https://tinyurl.com/5fpr56e4}}) to review it. If the revision is performed by a user with more than 2,000 reputation scores, the revision will be applied to a question without any review process. 
An overview of the ARQ revision process is presented in Figure \ref{ExampleOfQuestionRevisionThread}. An asker posted an ARQ asking about microservice authentication strategies. This question was asked on April 15, 2015, at 8:10. However, later on the same date, just after almost four hours (April 15, 2015, at 12:29), another user (i.e., commenter) was not satisfied with the information provided in this question because the question missed some important information (specifically, the relationship between authentication service and Redis in microservice authentication strategy in the question body) by stating that: “\textit{Would you please provide more details on what you are trying to achieve? In the first case does authentication happen against Redis, or in the services themselves? Redis is missing in the second diagram, is this intentional?}”. Then the asker revised the question and stated in the comment: “\textit{I have added some information. Please let me know if it is still unclear. Thanks}”. 

\section{Related Work} \label{relatedwork}
\subsection{Architectural Information in Q\&A Sites}
Mining architectural information from Q\&A sites has been explored by both researchers and practitioners \citep{jean2024mining}, and a number of existing studies have studied architectural information provided in ARPs on SO from different perspectives. \cite{bi2021mining} developed a semi-automatic approach to mine ARPs from SO and structured the design relationships between architectural tactics and quality attributes in practice to help architects better make design decisions. \cite{soliman2021exploring} conducted an empirical study with 50 software engineers, who used Google to make design decisions using the Attribute Driven Design steps \citep{cervantes2016designing}. Based on the relevance and Architecture Knowledge (AK) concepts specified by software engineers, they determined how effective web search engines are to find relevant architectural information from various sources (including Stack Overflow) and capture AK. In another work, \cite{soliman2017developing} developed an ontology that covers AK concepts in SO. The ontology provides a description of architecture information to represent and structure AK in SO. With the goal of improving how architects search for architecturally relevant information, \cite{soliman2018improving} developed a new search approach (i.e., a domain specific-search approach) for searching architecturally relevant information using SO. They found that the new search approach outperforms the conventional keyword-based search approach (searching through the search engines, such as Google). \cite{tian2019developers} conducted an empirical study of SO users’ perception of architectural smells by analyzing the discussions from architecture smell related posts in SO. They found that SO users often describe architectural smells with some general terms, such as “bad”, “wrong”, “brittle” or violation of architectural patterns. \cite{zou2017towards} used topic modeling to analyze non-functional requirements related to textual content in SO posts in order to understand the actual requirements of developers. \cite{chinnappan2021architectural} extracted data from five data sources, including Stack Overflow and source code repositories, and mined architectural tactics for energy-efficiency robotics software applied by practitioners in real robotics projects. To foster the applicability of the identified tactics (even beyond the robotics software community), they described them in a generic and implementation independent manner by UML component and sequence diagrams. The presented energy-aware tactics can serve as guidance for roboticists, as well as other developers interested in architecting and implementing energy-aware software. \cite{de2023characterizing} conducted an exploratory study of architecture information discussed on SO by analyzing ARPs shared on that site. They found that SO users ask a broad range of architecture-related questions, including questions about architecture configuration, architecture decision, architecture implementation, and architecture refactoring. They also found that SO users provide a wide range of architecture solutions (including architecture patterns, tactics) to those questions.

%Our work differs from the aforementioned work in that the existing work mainly investigated architectural information shared in ARPs in terms of the types of architecture knowledge discussed in SO posts (i.e., questions and answers), the perception of architecture concepts (e.g., architecture smell) by SO users, while our study focuses on the revision of information shared in ARQs. Moreover, prior work is based on ARQs and their associated answers, while our work covers the entire ARP, including its question and answers, comments under the question, and comments under the answers. 
\textcolor{black}{Our work differs from the aforementioned studies in that prior research primarily examined architectural information shared in ARPs by analyzing the types of architecture knowledge discussed in SO posts (i.e., questions and answers) and users’ perceptions of architectural concepts (e.g., architectural smells). In contrast, our study focuses on how information in ARQs is revised and how these revisions influence the answers, an aspect not addressed in previous studies. Furthermore, prior work typically analyzes both ARQs and their answers (often focusing on accepted answers) to identify types of architectural knowledge, and did not consider information shared in comment threads in their analysis. Our study broadens this scope by including comments (both on questions and answers) in addition to the ARQ and all its answers (regardless of whether they are accepted or not). This broader scope provides a distinct perspective and highlights the methodological differences between our work and earlier research.} 
 
\subsection{Revising Information in Q\&A Sites}
\cite{chen2017community} proposed a deep learning based approach for learning to apply sentence editing patterns from large-scale community post edits. The authors evaluated their approach through large-scale archival post edits and the approach demonstrated promising quality of recommended sentence edits. \citep{li2015good} examined the collaborative editing of posts (i.e., both answers and questions) on SO, and explored its benefits on content quality and potential negative effects on users activity. They found that collaborative editing could improve the number of positive votes, which implies an increase of the quality of posts. \cite{zhu2022empirical} investigated how question discussions (i.e., questions with their attached comments) are associated with the evolution of questions. Specifically, they studied the association between the number of comments and question revisions to better understand how question discussions affect the evolution of the question contents. They found that questions with chat rooms are more likely to be revised than questions without chat rooms, with a median size increase  of 114 characters. They also found that there is a strong correlation between the number of question comments and the question answering time (i.e., more discussed questions receive answers more slowly). \cite{naghashzadeh2021users} studied the evolution of MATLAB questions on SO, and they found that most revisions of MATLAB questions are text-related but not code snippets.

These studies are related to our work since they investigated the revisions of SO posts. Nevertheless, the purposes of the aforementioned work and our work are different. The abovementioned studies did not investigate the revision of SO posts, specifically, ARPs. To the best of our knowledge, there has been no investigation of the revision of information provided in ARPs (e.g., ARQs) with regard to types of the missing and further provided information in ARQs, purposes of the further provided information in ARQs, and impact of the further provided information in ARQs on the answers, which is the focus of this study. Our study complements the existing work on the revision of SO posts through analyzing ARPs. 
%Munteanu e\textit{t al}. presented a design of a webcast extension that engages users to collaborate in a Wiki-like manner on editing the transcripts that are produced by automatic speech recognition techniques \cite{munteanu2008collaborative}. Munteanu \textit{et al}. showed that this is a feasible solution to improve the quality of transcripts. Kittur et al. examined how the number of editors on Wikipedia and the coordination methods that they used affect the quality of Wikipedia article \cite{kittur2008harnessing}. They observed that adding more editors has no association with improvements in the quality of articles, especially when the work was distributed evenly among editors or when they used explicit communication on the article talk page to coordinate. Calvo \textit{et al}. proposed an architecture for supporting collaborative editing for academic writing \cite{calvo2010collaborative}. They analyzed the impact of writing activities on the quality of outcomes. These prior studies mainly focused on investigating the impact of collaborative editing on the quality of user generated contents (e.g., answer posts) and found that collaborative editing could improve the quality of articles in general. Our study focused on
\section{Methodology} \label{Methodology}
\begin{figure} 
 \centering
 \includegraphics[width=1.0\linewidth]{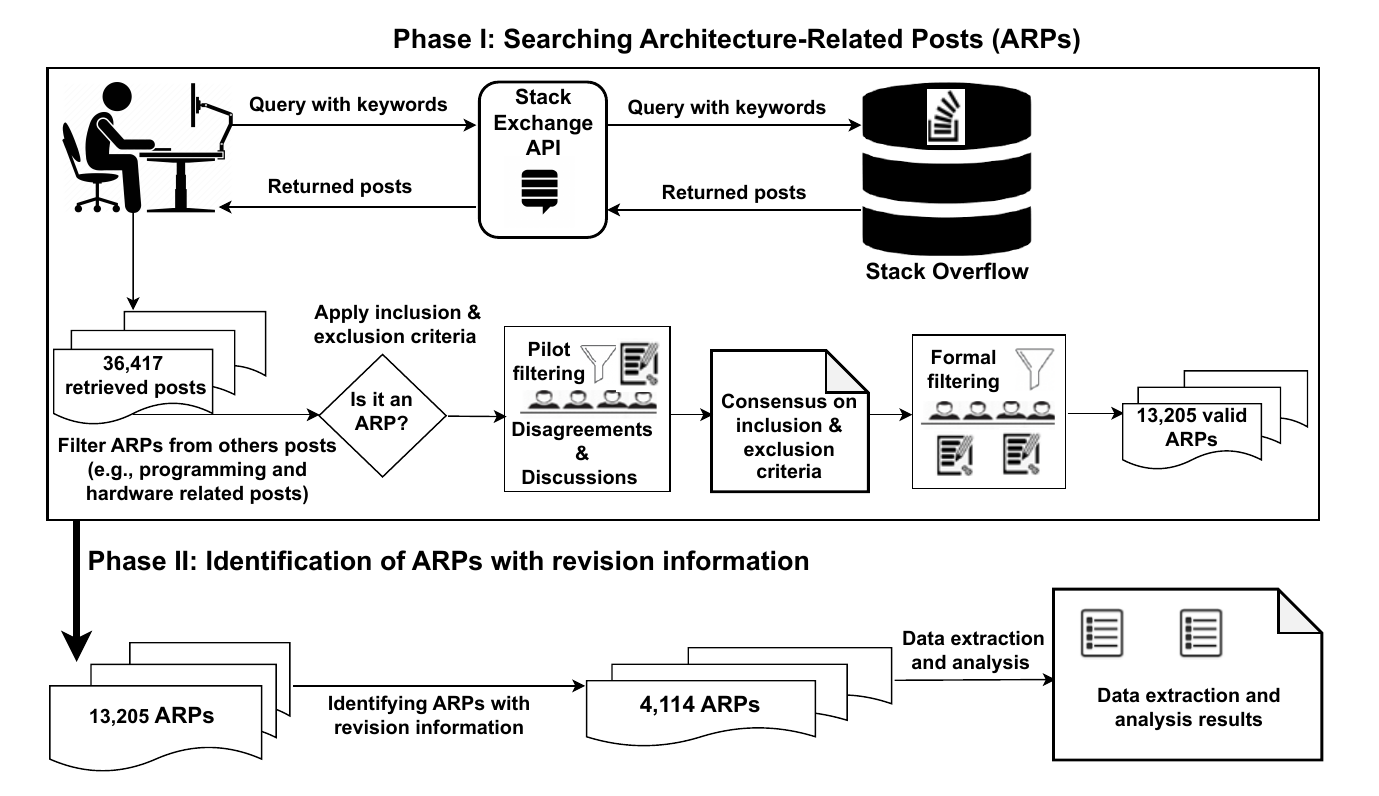}
 	\caption{Overview of the research process}
 \label{DrawingStudyExcution}
\end{figure}

The goal of this study is \textit{to \textbf{analyze} the current situation on revising ARQs \textbf{for the purpose} of understanding \textbf{with respect to} the prevalence of ARQ revisions, the types of the missing and further provided information in ARQs, the purposes of the further provided information in ARQs, and the impact of the further information provided in ARQs on the answers \textbf{from the point of view of} SO users \textbf{in the context of} software development in practice}. In the following subsections, we explain the Research Questions (RQs), their rationale, and the overview of the research process (see Figure \ref{DrawingStudyExcution}) used to answer the RQs. 

\subsection{Research Questions}\label{ResearchQuestions}

\textbf{RQ1. How \textcolor{black}{frequently} are architecture-related questions revised?}

\textbf{Rationale}: SO hosts a large number of problems or questions related to architecture design \citep{soliman2018improving, soliman2021exploring, de2023characterizing}, and these questions are asked in various domains or contexts \citep{de2023characterizing}. From the user’s point of view, creating an answer or architectural solution to an ARQ can be challenging since the initial version of an ARQ could be incomplete or ambiguous. For this reason, potential answerers may wish to engage the asker in a discussion to clarify the intent of the ARQ and possibly seek additional information, which is typically done using comments attached to the ARQ. If the discussion proves to be fruitful, the asker may decide to revise the original question to clarify the intent of this question to the answerers. Then, the answerers may post answers based on the discussion in the comments. However, we do not know how \textcolor{black}{frequently} ARQs are revised after being posted. By answering this RQ, we aim to explore the distribution and proportion of ARQ revisions, which can help get a basic overview of ARQ revisions on SO. 

\textbf{RQ2. What information is missing and further provided in architecture-related questions after being posted?}

\textbf{Rationale}: Asking ARQs on SO may not always be straightforward. For instance, ARQs may miss or lack descriptions or explanations for certain important architectural elements (e.g., architecture patterns \citep{SA2012} and design contexts \citep{bedjeti2017modeling}), which may, in turn, make such questions difficult to understand and consequently get few or no answers/solutions. By answering this RQ, we intend to identify and classify the frequently missing and further provided information in ARQs, which is beneficial for researchers and practitioners to understand and improve the way ARQs are asked and revised in SO.

\textbf{RQ3. What are the purposes of the further information provided in architecture-related questions?}

\textbf{Rationale}: SO users revise or provide further information in ARQs on SO for different purposes, such as clarifying the understanding of architecture under design. By answering this RQ, we intend to understand the main purposes of revising or providing further information in ARQs by SO users, which can facilitate the development of techniques and tools that support these purposes.

\textbf{RQ4. What is the impact of the further information provided in architecture-related questions on the answers?}

\textbf{Rationale}: The importance of high-quality content in SO, has been recognized in many studies. For example, Li \textit{et al}. observed that good answers are more likely to be given in response to good questions~\citep{li2012analyzing}. Revising ARQs can have not only a positive impact on the quality of ARQs, but also on the quality of answers/architecture solutions. By answering this RQ, we aim to understand the impact of revising or providing further information in ARQs on the architecture solutions, which could help provide suggestions for SO users to structure and compose ARQs with the likelihood of getting quality architecture solutions. 

\color{black}
\textbf{RQ5. How do practitioners perceive the identified categories related to the revision of architecture-related questions on Stack Overflow?}

\textbf{Rationale:} Understanding practitioners' perceptions is crucial to evaluating the quality and practicality of the identified categories in ARQ revisions. By answering this RQ, we aim to assess the quality and applicability of these categories from the perspective of experienced practitioners. Their expert feedback can both corroborate our classification schema and uncover any deficiencies, thereby guiding the development of effective tools and guidelines that enhance the formulation and refinement of ARQs on SO. 
\color{black}

\subsection{Data Collection and Filtering} \label{datacollectionAndFiltering}
To answer the RQs, we need to search for and collect the data related to the revisions of ARQs in SO. Our data collection and filtering process is divided into two phases, namely \textit{Phase I: Searching architecture-related posts} and \textit{Phase II: Identification of ARPs with revision information}, as detailed below:

\textit{\textbf{Phase I: Searching architecture-related posts}} \label{PhaseI}

\textit{a) Search terms}:\label{searchquery}
Before we decided on the most suitable terms for retrieving posts relevant to architecture design, we first performed a pilot search with several terms, namely “architect*” (i.e., “architect”, “architecture”, “architectural”, and “architecting”) and “design*” (i.e., “design” and “designing”), within SO. The process was carried out by using an SQL query through the query interface provided by StackExchange Data Explorer\footnote{\url{https://data.stackexchange.com/stackoverflow/query/new}}, which is a web interface that allows the execution of SQL queries on data from Q\&A sites, including SO. After the pilot search with the mentioned terms, we saw that SO users mostly use the terms “design*” (i.e., “design” and “designing”) in the programming context on SO, for example, singleton design pattern\footnote{\url{https://tinyurl.com/vaak5a3c}}. 

Therefore, we selected the terms “architect*” {(i.e., “architect”, “architecture”, “architectural”, and “architecting”)} to be used in our search. It is worth mentioning that we did not use the search terms to search exclusively through tags only because tags can sometimes be less informative and ineffective \citep{chen2019modeling}. There are several disadvantages to using tags as the only approach to determine whether a post is related to a topic. This is due to the reason that a user who created a post could be unsure about the title of the most appropriate tag for their discussion, which can lead to the use of incorrect or irrelevant tags \citep{chen2019modeling}. For example, in this ARP (SO post \#21588543) that asked for an architecture pattern that can be used in the design of a single webform application, a developer used tags (“jc\#”, “asp.net”, and “web”), and these tags cannot immediately tell in which contexts (e.g., architecture or programming context) they are really used. Another problem with user-defined tags is that users may try to add as many tags as possible (SO allows up to 5 tags) to raise the number of views and probably increase the probability of getting responses quickly \citep{chen2019modeling}. Thus, while tags can be helpful to capture posts related to architecture design, using tags exclusively may miss important posts on this topic. Hence, we decided to add the title and body of the questions into the search. In our replication package~\citep{dataset}, we provided the SQL query that we used to search ARPs in SO, such as how tags, title, and body of a post were combined during the search. The searching process resulted in 36,417 posts (see Figure \ref{DrawingStudyExcution}). 

\textit{b) Filtering ARPs from other posts (i.e., programming and hardware related posts)}:\label{FilteringARPs}
We found that SO users use the term “architecture” not only in the context of software architecture design but also in other contexts, such as hardware architecture context (e.g., ARM6 CPU architecture\footnote{\url{https://tinyurl.com/3jzfpr6e}}) and programming context (e.g., array architecture\footnote{\url{https://tinyurl.com/2khrtykr}}), when describing their problems in the SO posts. Therefore, we need to filter the retrieved 36,417 posts and exclude those posts related to programming and hardware architecture. To achieve this, we performed context analysis and employed the defined inclusion and exclusion criteria (see Table \ref{mainInclusionExclusion}) to dedicatedly filter and separate software ARPs from other types of posts mentioned above.
 
\textcolor{black}{Before the formal post filtering (manual inspection), a pilot filtering was conducted to establish a shared understanding of the inclusion and exclusion criteria (see Table \ref{mainInclusionExclusion}). The first author randomly selected a sample of 1,000 posts from the 36,417 posts and manually applied the defined criteria. The second and third authors checked and examined the filtering results to ensure consistency and alignment in the interpretation of the criteria. Through this process, the first three authors reached a consensus on the application of the inclusion and exclusion rules. During this pilot filtering, 51 posts led to disagreements or misunderstandings. These cases were discussed and resolved among the first three authors using a negotiated agreement approach \citep{campbell2013coding}. Following the pilot phase, the first author proceeded with the full manual filtering. To manage the large dataset efficiently, the remaining posts were divided into manageable subsets of 1,000, and each subset was reviewed sequentially. This process continued until all 36,417 posts were manually inspected, resulting in 13,205 candidate ARPs (see Figure \ref{DrawingStudyExcution}). The final results were reviewed and verified by the other authors. This step took twenty-one full days (approximately 12 hours per day) to complete, during which we identified and separated ARPs from programming and hardware architecture posts. At this stage of formal filtering, no further disagreements or misunderstandings were encountered.}

\begin{table*}[h!]
\small
\footnotesize
\caption{Inclusion and exclusion criteria for filtering ARPs}
\label{mainInclusionExclusion}
\begin{tabular}{m{11.5cm}} 
\hline
 \textbf{Inclusion criteria}                          \\ \hline
                                                               
\textbf{I1}. A post contains a discussion on software architecture, for example, architecture design and architecture pattern. \\ 
\textbf{I2}. A post contains at least one answer attached to its question as we aim to study the impact of the further provided information in an ARQ on its answer(s).\\ \hline

\textbf {Exclusion criteria}\\ 
\midrule
\textbf{E1.}  A post that has a score (i.e., medium number of down/upvote) less than 1 is excluded since we want to make sure that the studied posts have attracted enough attention from the community \citep{UnderstQuestQuali2014}.\\	 
\bottomrule 
\end{tabular}
\end{table*}
\normalsize
 
\textit{\textbf{Phase II: Identification of ARPs with revision information}}\label{PhaseII_ARPwithRevisionInfo}

For answering our RQs, we need to define a set of criteria and filter the 13,205 ARPs to identify ARPs with the revision information (i.e., ARPs with the questions that were revised by SO users). We observed that SO users often use terms related to revision, such as  “revis*” (i.e., “revise”, “revision”, “revising”, “revised”), “edit*” (i.e., “edit”, “edited”, “editing”), and “updat*” (i.e., “update”, “updated”, “updating”) in the bodies of the questions to refer to the revision of ARQs in SO (see Figure \ref{ExampleOfQuestionRevisionThread}). Also, we observed that SO users occasionally use similar terms related to revision (e.g., “edit”, “edited”, “editing”, “update”, “updated”, “updating”) in the comment threads to refer to the revision of ARQs. Moreover, the studies by \cite{obsolete2019} and \cite{readingAnswers2019} argued that the quality of a post (e.g., answer) is a combination of both the answer and its associated comments as comments may provide additional information to support the answer, such as improvement of answers \citep{readingAnswers2019} and obsoleted answers \citep{obsolete2019}. Therefore, in this study, we included the information in comments (i.e., all comments posted under ARQs and answers) to gain a deep understanding of how SO users perform the revisions of ARQs and the impacts of these revisions on the answers on SO. Thus, based on this observation along with the aid of our defined selection criteria in Table \ref{CriteriaForARPsWithRevisionInfo}, we manually checked and filtered the 13,205 ARPs to identify ARPs with revision information (each ARP includes a question with its answers and all the comments that are associated with the question and answers).

Specifically, to filter out ARPs that do not discuss the revisions of ARQs from the 13,205 ARPs and reach an agreement about the criteria defined in Table \ref{CriteriaForARPsWithRevisionInfo}, we first executed a pilot ARP filtering. During the pilot post filtering, the first and second authors selected a random dataset of 10 ARPs and checked them independently with the help of the criteria defined in Table \ref{CriteriaForARPsWithRevisionInfo}. To measure the inter-ratter agreement between the first two authors, we calculated the Cohen’s Kappa coefficient \citep{cohen1960coefficient} of the pilot post filtering and got an agreement of 0.912. Then, the first author carried on to check and filter the remaining ARPs. For the posts that were unclear and the first author got confused while filtering, physical meetings with the second author were scheduled to solve such confusion. The number of resulting ARPs (with revision information) that were used to answer RQs is 4,114 (see Figure \ref{DrawingStudyExcution}).

\begin{table*}[h!]
\small
\footnotesize
\caption{Inclusion and exclusion criteria for identifying ARPs with revision information}
\label{CriteriaForARPsWithRevisionInfo}
\begin{tabular}{m{11.5cm}} 
\hline
 \textbf{Inclusion criteria}                          \\ \hline
                                                               
\textbf{I1}. A comment posted under an ARQ contains one of the keywords related to revision (such as “revise”, “revising”, “revised”, “edit”, 
       “editing”, “edited”, “update”, “updating”, and “updated”) and this comment is used to signify the revision of the ARQ. \\ 
\textbf{I2}. The title or body of an ARQ is revised/edited/updated.\\ \hline

\textbf {Exclusion criteria}\\ 
\midrule
\textbf{E1.} The keyword related to the revision in an ARQ (such as “edit”, “editing”, “edited”, “update”) is used to talk about something else (e.g., an ARQ itself is related to a “revision” topic rather than being a sign that the ARQ was required to be revised) other than the revision of the ARQ.\\
	 
\textbf{E2.} An ARP with conflicting discussions on the revision of the ARQ (e.g., if there are two comments posted under the ARQ and one comment states the need for revising the ARQ while another states no need) is not included.\\
\bottomrule 
\end{tabular}
\end{table*}
\normalsize

\subsection{Data Extraction and Analysis} \label{dataExtractionAndAnaly} 

\subsubsection{Data Extraction} 
To answer the defined RQs of this study (see Section \ref{ResearchQuestions}), we carefully read all the ARPs selected in this study (i.e., 4,114 ARPs) (see Figure \ref{DrawingStudyExcution}) and extracted the required data items as listed in Table \ref{DataExtraction}. During this process, the first author conducted a pilot data extraction with 10 ARPs independently, and any unsure extraction results were discussed with the second author until they reached a consensus to increase the correctness of the extracted data. The first author further extracted data according to the data items from the rest selected ARPs, marked the uncertain sentences or paragraphs, and then discussed them with the second author. The first author reexamined the extraction results of all the ARPs to make sure that all the data were extracted correctly. Table \ref{DataExtraction} also shows the RQs that are supposed to be answered using the extracted data. The data extraction was subsequently followed by data analysis, and these two processes were conducted and recorded with the aid of MAXQDA (a qualitative data analysis tool). 

To answer RQ1, we used all the ARPs (i.e., 13,205) after Phase I(b) of the data collection and filtering process (see Section \ref{FilteringARPs}) as the raw data, and we counted how many ARPs in the raw data contain revision information (which resulted in 4,114 ARPs) to investigate the prevalence of ARQ revisions on SO. To answer RQ2, RQ3, and RQ4, we analyzed all the ARPs that contain revision information obtained from RQ1 (i.e., the ARPs after Phase II, see Section \ref{PhaseII_ARPwithRevisionInfo}) to get the missing and further provided information in ARQs, purposes of the further provided information in ARQs, and impacts of the further provided information in ARQs on the answers.

\begin{table*}
\small
\footnotesize
\caption{Extracted data items and their corresponding RQs}
\label{DataExtraction}
\begin{tabular}{m{0.1cm}m{4cm}m{5cm}m{1.1cm}<{\centering}} 
\hline
\textbf{\#}                                             & \textbf{Data item}       & \textbf{Description}        &\textbf{RQ} \\ \hline
 D1                                                     & Existence of an ARQ revision %Selected 
                                                        & Whether or not an ARQ contains revision information 
                                                        & RQ1\\\cline{1-4}
                                                        
 D2                                                     & Missing and further provided information in an ARQ  
                                                        & Type of information that was missed in the first version of an ARQ 
                                                        & RQ2\\\cline{1-4}
                                                        
 D3                                                     & Purpose of the further provided information in an ARQ 
                                                        & The intention of proving further information in an ARQ 
                                                        & RQ3\\\cline{1-4}

 D4                                                     & Impact of the further provided information in an ARQ on the answers 
                                                        & The effect of the further provided information in an ARQ on the Answer
                                                        & RQ4\\\cline{1-4}                                                      

 \textcolor{black}{D5}                                   & \textcolor{black}{Practitioners' perceptions of the categorization of ARQ revisions}  
                                                        & \textcolor{black}{Evaluate the quality of the identified categories on the revisions of ARQs}  
                                                        & \textcolor{black}{RQ5}\\\cline{1-4}

\end{tabular}
\end{table*}
\normalsize

\subsubsection{Data Analysis} \label{dataAnalyis}
In this section, we analyzed the extracted data (see Table~\ref{DataExtraction}) to answer the research questions (RQs) of this study. \textcolor{black}{Following prior work~\citep{wu2019developers}, we applied open coding and constant comparison as standalone qualitative analysis techniques to answer RQ2, RQ3, and RQ4, guided by recommendations from Seaman~\citep{seaman1999qualitative}. Descriptive statistics~\citep{wohlin2003empirical} were used to answer RQ1. We note that open coding and constant comparison were employed independently, rather than as part of the full grounded theory methodology. While these methods are foundational to grounded theory and often support theory development, we did not extend our analysis to later phases such as axial or selective coding. This decision was guided by three main reasons: (1) the descriptive and exploratory nature of our research questions; (2) the objective of our study, which is to explore and categorize recurring themes in the data rather than to develop a core category or propose a theoretical model, as is typically the goal in grounded theory; and (3) the scope and structure of our dataset, which did not necessitate the complexity of a full grounded theory approach. Therefore, open coding and constant comparison offered sufficient methodological rigor for identifying meaningful patterns without engaging in the additional coding stages characteristic of grounded theory.}

\textbf{RQ1: How \textcolor{black}{frequently} are architecture-related questions revised?}

We used descriptive statistics \citep{wohlin2003empirical} to answer RQ1. Specifically, we investigated this RQ in three aspects: (1) the percentage of ARQs that contain revision information, (2) the ARQ revisions by events, namely the ARQ revision before and after the question receives its first answer, and (3) the percentages of ARQ revisions by two types of users, namely Question Creators (QCs) and other users, i.e., Non-Question Creators (N-QCs). %(4) the proposition of the ARQ revisions across different application domains. 
By analyzing these three aspects, we could get an overview of how often ARQs are revised, as well as who and when revise the ARQs. The data analysis approaches for these three aspects are detailed below:

\textit{(1) The proportion of ARQs that contain revision information}. We analyzed the percentage of ARQs that contain revision information in the raw data. Specifically, we calculated the percentage of 4,114 ARQs on 13,205 ARQs, and the details about the percentage are provided in Section \ref{ResultsOfRQ1}.  

\begin{figure} 
 \centering
 \includegraphics[width=1.0\linewidth]{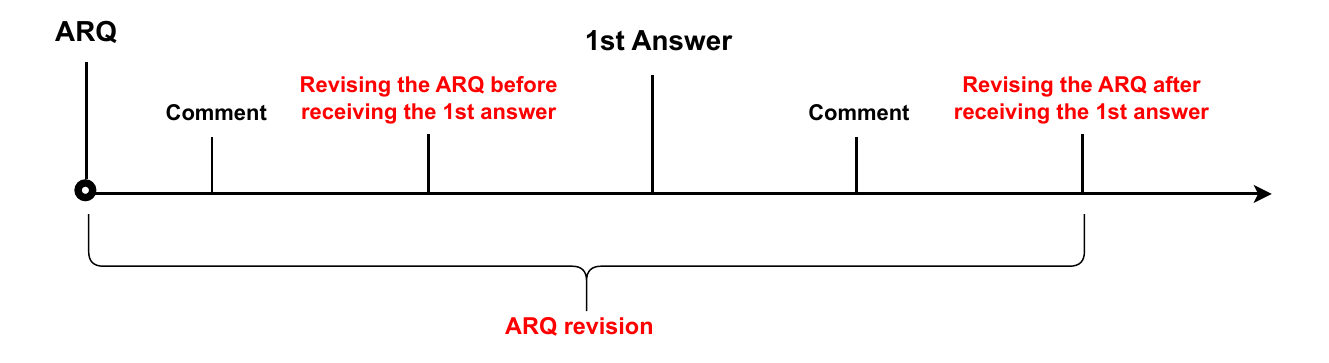}
 	\caption{Timeline of ARQ revision events (ARQ revision can occur at any time since the creation of a question)}
 \label{TimelineForQuestion}
\end{figure}

\textit{(3) The ARQ revisions by events}. As depicted in Figure \ref{TimelineForQuestion}, after an ARQ is posted on SO, several different types of follow-up events may occur\footnote{\url{https://tinyurl.com/5e956szz}}. For example, after an ARQ is posted, any of the following events can occur: users can post comments to discuss the question, the ARQ can be edited for clarity, users can propose answers to the ARQ, and users can still revise the question for more clarity. We studied when an ARQ revision occurs relative to two key events in the Q\&A process, namely before and after the ARQ receives its first answer/architecture solution. For each ARQ, we constructed the timeline consisting of each event (see Figure \ref{TimelineForQuestion}), and we analyzed the prevalence of ARQ revisions with respect to those events. Note that the ARQ on SO can be revised more than one time and from the 4,114 ARPs (i.e., 4,114 ARQs), we found 8,794 corresponding revisions. Subsequently, we counted the revisions made in each ARQ before and after the question received its first architecture solution. 

\textit{(2) The proportion of ARQ revisions by users}. We categorized SO users into two groups: Question Creators (QCs) and Non-Question Creators (N-QCs). QCs are the users who asked or posted ARQs, whereas N-QCs are other users (such as commenters and answerers) who helped revise ARQs. Our primary motivation was to understand the participation level of each user group in ARQ revisions.

\color{black}

\textbf{RQ2. What information is missing and further provided in architecture-related questions after being posted?}

\textcolor{black}{To answer RQ2, we applied open coding and constant comparison following qualitative analysis guidelines~\citep{seaman1999qualitative}. Specifically, during the open coding phase, the first author carefully examined the contents of the ARQs and their associated comments to identify units of meaning related to missing and further provided information. Each identified unit was assigned a descriptive code that summarized the key idea. For instance, in one ARQ (SO post \#7588037), the asker updated/edited the question to clarify an issue regarding module inter-dependencies. The updated text reads:}

\begin{quote}
\textcolor{black}{\textit{``Update/Edit: There are two places where my modules are inter-dependent. Following the approach discussed below, I can move the interfaces and types I want shared from Module1 to Core which allows ModuleB to make use of them without tightly coupling ModuleA and ModuleB. However, I also need to use some of the data contracts defined in ModuleA.Api for the service operations in ModuleB.Api. This makes me question the design (...).''}}
\end{quote}

\textcolor{black}{This edit was initially coded as ``module inter-dependency''. Next, during the constant comparison phase, the first author compared new codes with existing ones to identify similarities and differences. Related codes were grouped into broader concepts. For example, the code ``module inter-dependency'' was grouped under the broader concept ``module dependency.'' Subsequently, similar concepts were further abstracted and organized into higher-level categories; in this case, ``module dependency'' contributed to the broader category ``component dependency'' (see Table~\ref{CategoriesOfMissingAndFurtherProvidedInfo}). To mitigate potential bias and enhance the reliability of the analysis, the second and third authors independently checked and validated the generated codes, concepts, and categories. Discrepancies were discussed and resolved through a negotiated agreement approach~\citep{campbell2013coding}. This collaborative validation helped ensure that the final categories were meaningful, reliable, accurate, and reflective of the data.} As a result of this iterative process, we derived 14 categories capturing the types of missing and further provided information in ARQs, which are detailed in Section~\ref{ResultsOfRQ2}.

\textbf{RQ3. What are the purposes of the further information provided in architecture-related questions?}

\textcolor{black}{Similar to the analysis conducted for RQ2, we employed open coding and constant comparison to answer RQ3. We observed that SO users often mentioned the purposes of the further provided information either in the bodies of the ARQs or in the comments attached to them. 
For example, under the ARQ from SO post \#3993121, two users posted comments: one user remarked ``\textit{What's the question? ‘Is this a good thing?’ You will probably not get any good answers to such an open-ended question}'', prompting the asker to reply. An asker replied that ``\textit{Sorry, I should have been more clear (edited now) (...) I'm just concerned as to how this platform/architecture will scale and how secure it will be. Thanks}''. In this case, the first author summarized these interactions and encoded them as ``concern with scalability and security of architecture''. The first author then followed the same analytical procedures used for RQ2, including grouping similar codes into higher-level concepts and categories. 
For example, the code ``concern with scalability and security of architecture'' was merged into the concept ``clarify the understanding of architecture concern'', which was categorized under the ``clarify the understanding of architecture under design'' category (see Table~\ref{CategoriesOfPurposesOfTheProvidedInfo}). To mitigate potential personal bias, the second and third authors participated in checking and validating the generated concepts and categories. In cases of disagreement, we held meetings and used the negotiated agreement approach~\citep{campbell2013coding} to reach consensus, thereby improving the reliability of the analysis results for RQ3.} As a result of this analysis, we derived four categories of purposes for the further provided information in ARQs, which are fully elaborated in Section~\ref{ResultsOfRQ3}.

\textbf{RQ4: What is the impact of the further information provided in architecture-related questions on the answers?}

\textcolor{black}{SO provides mechanisms for revising posts (i.e., questions and answers)\footnote{\url{https://stackoverflow.com/help/privileges/edit}}. With these mechanisms, SO users can track the revisions made in the posts. For example, SO users can check and see what information (e.g., design context of an ARQ) has been added or revised in the question. In addition, SO users can inspect and see the information, such as when (the timestamp of) a question was posted, revised, or got an answer (see Figure~\ref{ExampleOfQuestionRevisionThread}). By checking the timestamps (such as the timestamp when the question was posted, the timestamp when the question was revised, or the timestamp when the question got an answer), it is clear to see and determine whether adding or revising certain information in a question has an impact on the answers or the question itself.}

\textcolor{black}{Therefore, when investigating RQ4, we gathered the following information from the 4,114 ARPs: the timestamp when an ARQ was posted, the timestamp when this ARQ was revised, and the timestamp when this ARQ got an answer after its revision. Moreover, we divided the study period of the answer into two stages according to the revision(s) made into its associated ARQ by examining the content of each answer and its attached comments before and after the revision(s) of the ARQ. Thereafter, we employed open coding \& constant comparison and referred to the gathered information from the 4,114 ARPs to investigate the impact of the further provided information in ARQs on the answers/architecture solutions.}

\textcolor{black}{For example, the ARQ (SO post \#8891979), which asked about the \textit{reusability of C\#-project on iPad}, was posted on Jan 17, 2012 at 9:07 and got updated/revised on Jan 17, 2012 at 9:53. In addition, this ARQ got the answer on Jan 17, 2012 at 9:40, which was edited on Jan 17, 2012 at 15:55, according to the revision made in the question on Jan 17, 2012 at 9:53. The update in the answer stated that ``\textit{Following your Update(2) in question: Yes, you can use a third party server to translate XAML-WPF-files to HTML5 - the ComponentArt Dashboard Server. This claims to translate WPF/Silverlight applications written using strict MVVM to HTML5/JS for portability across multiple devices}''. After four days (Jan 31, 2012 at 12:48) from the update time of the answer, the SO user who asked the question commented under this answer by stating that ``\textit{As far as I can see, the ‘ComponentArt Dashboard Server’ only works for the Controls, that are included in the ‘Framework’. I would have needed it for my own created controls. But nevertheless, I will accept that I have to rewrite it and you (with your answer) were very helpful! I like your inputs and indeed, the explanations are practically useful. Thanks}''.}

\textcolor{black}{In this case, the first author picked a phrase (i.e., a summary of that comment) and encoded that comment as ``ComponentArt Dashboard Server was practically helpful''. Subsequently, the first author went on to scrutinize the content of the answer (before and after the revision of the ARQ) under which this comment was posted to examine whether the impact on the answer (e.g., the answer/architecture solution becomes useful) is in fact due to the change/revision made in the ARQ. It is worth noting that the first author also examined what the change made in the ARQ was and whether the change is architecturally relevant. Then, the first author compared and grouped similar codes into concepts and categories as in the analysis of the previous RQs.}

\textcolor{black}{For example, the code ``ComponentArt Dashboard Server was practically helpful'' was merged into the concept ``make an architecture solution practically helpful'', which was regarded as the ``make an architecture solution practically useful/helpful'' category (see Table~\ref{CategoriesOfImpactsOfTheFurtherProvidedInfo}). The personal bias was mitigated through the validation of the generated codes, concepts, and categories with the other two authors (the second and third authors) of this study. Also, as in the analysis of the previous RQs, the disagreements were discussed and resolved in a meeting using the negotiated agreement approach~\citep{campbell2013coding} to improve the reliability of the analysis results for RQ4. After the final analysis, we obtained four categories of the impact of the further provided information in ARQs on the answers/architecture solutions. The details about these four categories for RQ4 are provided in Section~\ref{ResultsOfRQ3}.}

\color{black}
\subsection{Interview Study} 

To answer RQ5, we conducted interviews with experienced practitioners to evaluate the categories identified in our study. We describe the interview design, including participant recruitment and evaluation procedures in this section.

\subsubsection{Interview Design} 
We designed and conducted semi-structured interviews following the guidelines proposed by~\cite{personal2005}.

\subsubsection{Interview Structure}
When designing the interview questionnaire, we ensured that the interview Questions (Qs) were directly aligned with our main RQs (see the interview questionnaire in Table \ref{tab_interview_questions}). The interview begins with an introduction page that explains its purpose, provides preliminary definitions, and specifies the estimated completion time (i.e., 15 to 20 minutes). The interview consists of two main sections. The first section contains General Questions (GQs) (i.e., GQ1–GQ5), gathering information such as the participant's role (e.g., architect) and years of experience in software development. The second section focuses on the revisions of ARQs (i.e., Q6–Q16). We further divided the second section into four subsections: (A) \textit{Evaluating the Categories of Missing and further Provided Information in ARQs after Being Posted}, (B) \textit{Evaluating the Categories of Purposes of the Further Provided Information in ARQs}, (C) \textit{Evaluating the Categories of Impact of the Further Information Provided in ARQs on Answers}, and (D) \textit{Final Reflection}. In each subsection, we presented the participants with relevant categories identified in our study and asked targeted questions. For example, in subsection (A) (“\textit{Evaluating the Categories of Missing and further Provided Information in ARQs after Being Posted}”), participants were asked whether the identified categories were meaningful, reliable, and helpful in understanding typical patterns of missing and further provided information in ARQs after being posted. Overall, the interview comprised open-ended questions and semi-closed questions. To ensure sufficient context, participants were presented with at least two examples of ARPs representing each category from our dataset before responding to the corresponding Qs. After reviewing the relevant ARPs to understand their context, participants evaluated the identified categories based on the Qs. Prior to the formal study, a pilot was conducted with two participants to refine the Qs and ensure clarity and relevance.

\subsubsection{Participant Recruitment} 
We recruited 11 practitioners from software companies in Germany, China, France, and Australia, all of whom often or sometimes engaged in architectural discussions or sought architecture solutions on SO. On average, participants had four years of experience in software development and a solid understanding of architectural design concepts.

\subsubsection{Interview Data Analysis}
All interview sessions were audio-recorded and transcribed verbatim. Additionally, participants were given the opportunity to review and supplement their responses via the interview questionnaire (see Table \ref{tab_interview_questions}) after the session, allowing them to include any thoughts they may have omitted during the live discussion. We analyzed the qualitative responses using open coding and constant comparison. This process followed the same procedures used in the analysis of previous RQs (e.g., coding extracted data items and grouping similar codes into higher-level concepts and categories) to address RQ5. 

During the interview data analysis, we observed that several participant responses suggested new categories that could complement our existing categorization scheme. However, most of these proposed categories either aligned with or could be mapped to those identified categories in our exploratory study. For example, in response to Q8, one participant (P11) stated: \faHandORight \hspace{0.5mm} “\textit{From what I’ve seen, the existing categories are quite robust, but one possible addition could be something like 'deployment environment'. Sometimes, questions get revised to include where or how the architecture is intended to run, like cloud vs. on-premise, or specific infrastructure limitations, which can significantly affect the type of answer needed (...)}”. The first author summarized and coded this response as “architectural deployment context”, which was then grouped under the higher-level concept “missing contextual specification”. Through constant comparison, similar concepts were identified and merged into the overarching category “design context” (see Table~\ref{CategoriesOfMissingAndFurtherProvidedInfo}). To complement the qualitative analysis, we also applied descriptive statistics~\citep{wohlin2003empirical} to summarize the frequency and distribution of responses. The interview results are presented in Section~\ref{ResultsOfRQ5}, where we also describe how newly identified categories from the interviews were integrated with the findings from the exploratory study.

\textcolor{black}{The dataset collected and used in this study and the details of data analysis (e.g., coding in MAXQDA\footnote{\url{https://www.maxqda.com/}}) are available online for replication and validation purposes~\citep{dataset}.}

%####################################################################################################
{\small
\begin{longtable}{|>{\raggedright\arraybackslash}p{4cm}|>{\raggedright\arraybackslash}p{7cm}|}
\caption{Interview questions for evaluating the identified categories} \label{tab_interview_questions} \\
\hline
\textbf{Section} & \textbf{Interview Questions} \\
\hline
\endfirsthead

\multicolumn{2}{c}%
{{\bfseries Table \thetable\ (continued): Interview Questionnaire}} \\
\hline
\textbf{Section} & \textbf{Interview Questions} \\
\hline
\endhead

\hline
\endfoot

\hline
\endlastfoot

\multirow{5}{=}{\textbf{General Introduction}} 
& \textbf{GQ1}. Which country are you working in? \\
& \textbf{GQ2}. What is your primary role in software development? \\
& \textbf{GQ3}. How many years of experience do you have in software development? \\
& \textbf{GQ4}. What is the primary domain or industry of your organization? \\
& \textbf{GQ5}. How frequently do you use Stack Overflow and other Q\&A sites to do architecture tasks? \\
\hline

\multirow{3}{=}{\textbf{A. Evaluating the Categories of Missing and Further Provided Information in ARQs after Being Posted}} 
& \textbf{Q6}. Are the provided categories meaningful, reliable, and helpful in understanding what is typically missing and further provided in ARQs after being posted? \\
& \textbf{Q7}. Do the provided categories comprehensively capture all categories of missing and further provided information in ARQs after being posted? \\
& \textbf{Q8}. Are there additional categories of missing and further provided information in ARQs that should be included? \\
\hline

\multirow{3}{=}{\textbf{B. Evaluating the Categories of Purposes of the Further Provided Information in ARQs}} 
& \textbf{Q9}. Are the provided categories meaningful, reliable, and helpful in understanding the purposes of the further provided information in ARQs after being posted? \\
& \textbf{Q10}. Do the provided categories comprehensively capture all categories of purposes of the further provided information in ARQs after being posted? \\
& \textbf{Q11}. Are there additional categories of purposes of the further provided information in ARQs after being posted that should be included? \\
\hline

\multirow{4}{=}{\textbf{C. Evaluating the Categories of Impact of the Further Information Provided in ARQs on Answers}} 
& \textbf{Q12}. Are the provided categories meaningful, reliable, and helpful in understanding how ARQ revisions impact answers? \\
& \textbf{Q13}. Do the provided categories comprehensively capture all categories of impact of the further provided information in ARQs after being posted? \\
& \textbf{Q14}. Are there additional categories of impact of the further provided information in ARQs after being posted that should be included? \\
& \textbf{Q15} Based on your experience, do revisions in ARQs improve the relevancy, practicality, completeness, and informativeness of answers on Stack Overflow? \\
\hline

\textbf{Practical Relevance and Applicability of the Findings to Developers and Moderators} 
& \textbf{Q16}. Could these findings help practitioners (e.g., developers, Stack Overflow moderators) in improving the quality of ARQs? \\

\end{longtable}
}
%##################################################################################################
\color{black}

\section {Results}\label{Results}
%The results of this study are elaborated in this section. The prevalence of ARQ revisions (answer to RQ1) is presented in Section \ref{ResultsOfRQ1}. The missing and further provided information in ARQs (answer to RQ2) is described in Section \ref{ResultsOfRQ2}. The purposes of the further provided information (answer to RQ3) are presented in Section \ref{ResultsOfRQ3}. The impact of the further provided information in ARQs on the answers (answer to RQ4) is articulated in Section \ref{ResultsOfRQ4}. 

\subsection{Prevalence of ARQ Revisions (RQ1)}\label{ResultsOfRQ1}
To answer RQ1, we used descriptive statistics (as described in Section \ref{dataAnalyis}) to investigate (1) the proportion of ARQs that contain revision information, (2) the proportions of ARQ revisions by two types of users, namely Question Creators (QCs) and other users, i.e., Non-Question Creators (N-QCs), and (3) the proportions of ARQ revisions by events, namely the ARQ revision before and after the question receives its first answer/architecture solution.

\begin{figure}[h] 
 \centering
 \includegraphics[width=0.7\linewidth]{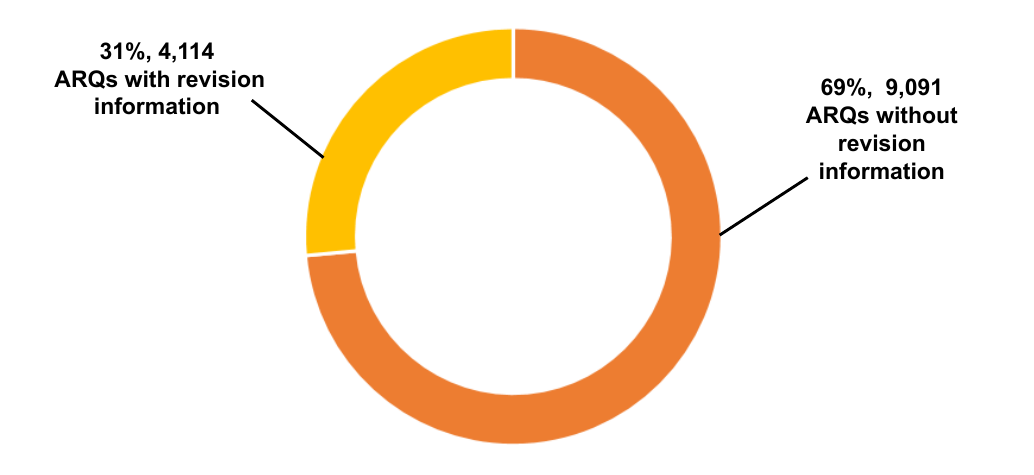}
 	\caption{The proportion of ARQs with and without revision information}
 \label{CountsPercentagesOfARQWithRevionsInfo}
\end{figure}

\begin{figure} 
 \centering
 \includegraphics[width=1.0\linewidth]{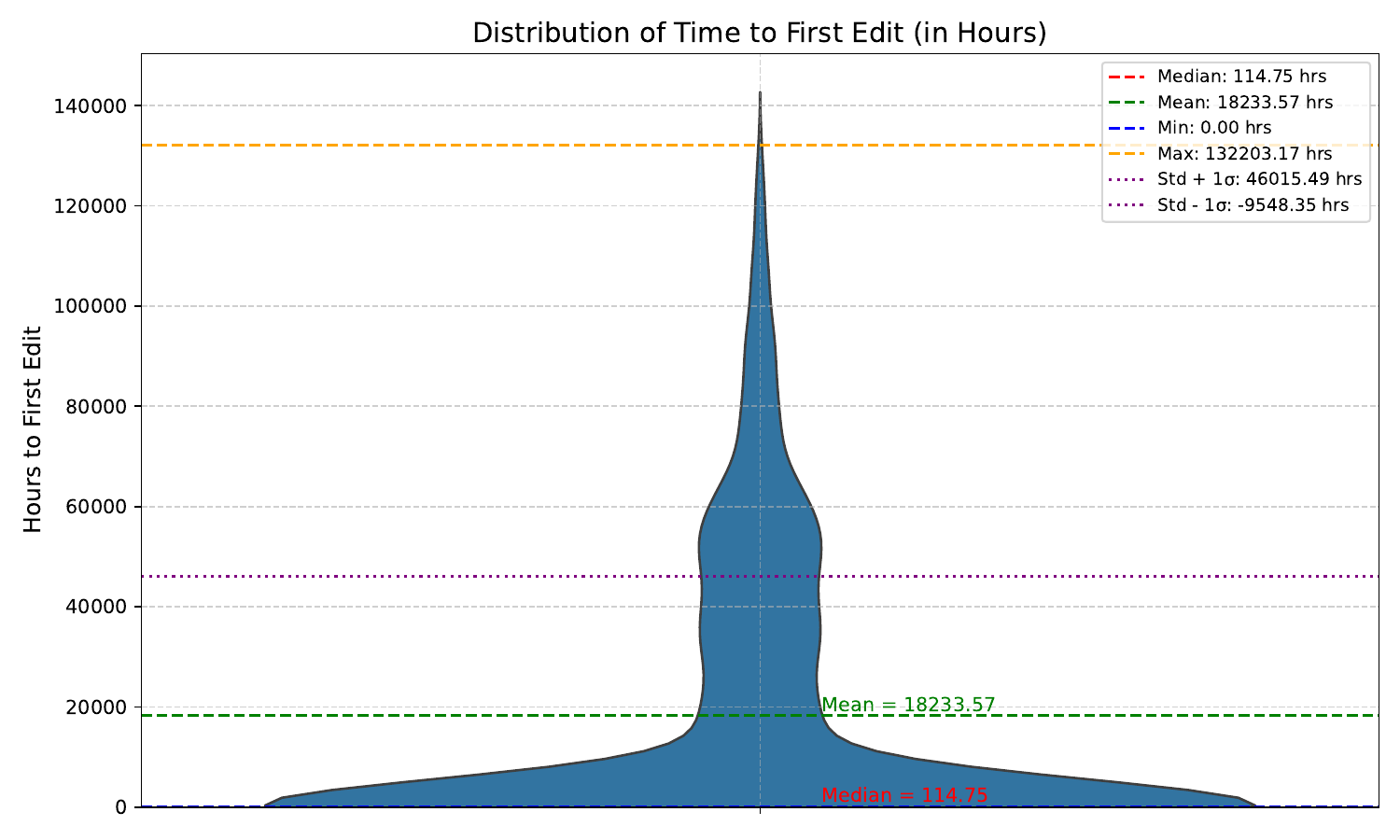}
 	\caption{\color{black}The distribution of time to first revision for ARQs in our dataset}
 \label{Violin_Plot_of_the_time_to_first_Revision}
\end{figure}

\textbf{The proportion of ARQs that contain revision information}: Figure~\ref{CountsPercentagesOfARQWithRevionsInfo} presents the counts and percentages of ARQs that contain revision information in our dataset. In total, we collected 13,205 ARQs from SO, and 4,114 of them contain revision information. Overall, the percentage of ARQs with revision information is 31\% only, which means that the revision of ARQs is not prevalent on SO. \textcolor{black}{Moreover, we examined the typical timeline for revisions, focusing specifically on how long it takes for the first edit to occur. Figure \ref{Violin_Plot_of_the_time_to_first_Revision} presents the distribution of the Time to First Edit (in hours) across all revised ARQs in our dataset, along with the following descriptive statistics: Minimum = 0.00 hours, Median = 114.75 hours, Mean = 18,233.57 hours, Maximum = 132,203.17 hours, and Standard Deviation = 27,781.92 hours. These results indicate that while some edits are made immediately after the question is posted, as reflected by the Minimum time of 0.00 hours, the Median time of approximately 115 hours (4.79 days) suggests that revisions typically occur within the first few days. In contrast, the Mean time is substantially higher due to a long tail of delayed edits. This skewed distribution is further illustrated by the violin plot, which highlights both the prevalence of early revisions and the occurrence of infrequent but significantly delayed edits.
%Moreover, we studied the typical timeline, specifically how long it takes for revisions to be done. Figure \ref{Violin_Plot_of_the_time_to_first_Revision} illustrates the distribution of the \textit{Time to First Edit} (in hours) across all revised ARQs in our dataset and accompanied by the following descriptive statistics: \textbf{Minimum} time to first edit: 0.00 hours, \textbf{Median} time to first edit: 114.75 hours, \textbf{Mean} time to first edit: 18,233.57 hours, \textbf{Maximum} time to first edit: 132,203.17 hours, \textbf{Standard deviation}: 27,781.92 hours. These results show that while the \textbf{Median} time to first edit is less than 2 hours, indicating that many edits occur shortly after the question is posted, the \textbf{Mean} time is considerably higher due to a long tail of late revisions. This pattern is clearly illustrated by the violin plot, which reveals the presence of both early revisions and infrequent but much later edits.
}

\color{black}

\begin{figure}[h]
   \centering
   \subfloat[The proportion of ARQ revisions\\by events]{\includegraphics[width=.5\linewidth]{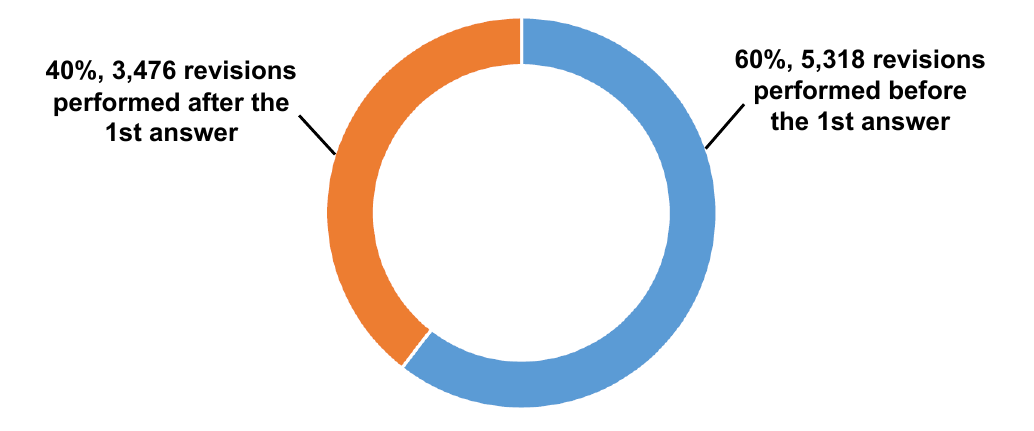}}\hfill 
   \subfloat[The proportion of ARQ revisions\\by users]{\includegraphics[width=.5\linewidth]{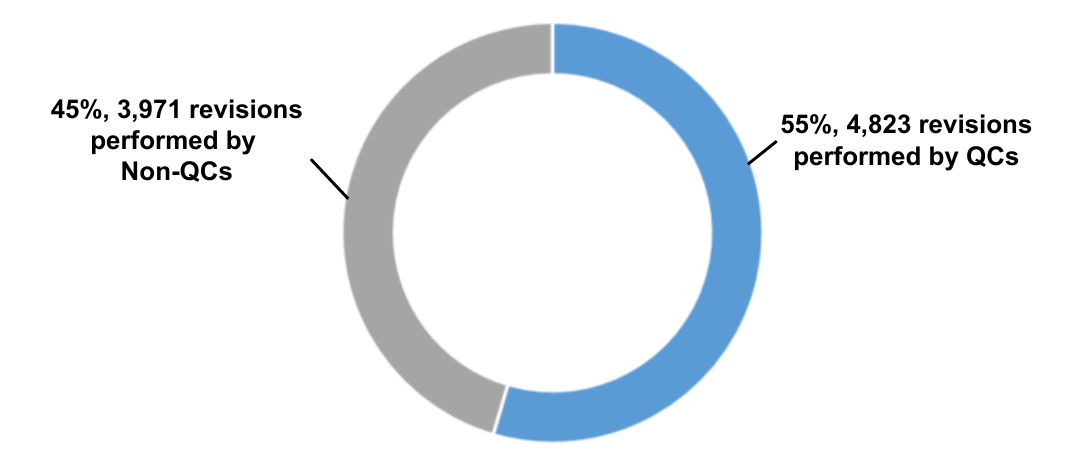}}\hfill
   \caption{The proportions of ARQ revisions by events and users}
   \label{PrevalenceOfARQRevisions}
\end{figure}

\textbf{The proportion of ARQ revisions by events}: As stated in Section \ref{dataAnalyis}, we studied when an ARQ revision occurs relative to two key events in the Q\&A process, namely before and after the ARQ receives its first answer/architecture solution. Moreover, we found that the ARQ can be revised more than one time and from the 4,114 ARPs (e.g., 4,114 ARQs) with revision information, we found 8,794 corresponding revisions. Subsequently, we counted the revisions made in each ARQ before and after the question received its first architecture solution. Figure \ref{PrevalenceOfARQRevisions}(a) shows the proportions of ARQ revisions relative to two key events earlier mentioned. We observed that an ARQ revision can occur at any time since the creation of this question. Specifically, the ARQ revision starts soon after the ARQ is posted (e.g., from 1 minute onward), with most ARQ revisions (60\%, 5,318 out of 8,794 revisions) occurring before the first answer/architecture solution is proposed and posted, and the ARQ revision continues even after the ARQ received the first architecture solution.

\textbf{The proportion of ARQ revisions by users}: Figure~\ref{PrevalenceOfARQRevisions}(b) depicts the proportions of users (QCs and non-QCs) who had revised ARQs according to the dataset we collected from SO. From the results, we can find that both QCs (55\%, 4,823 out of 8,794 revisions) and non-QCs (45\%, 3,971 out of 8,794 revisions) actively participate in ARQ revisions, with most ARQ revisions being made by the QCs. %It is not surprising that QCs are the primary users that revise their ARQs because revising or correcting the ARQs may require deep knowledge of the ARQs, such as design contexts and relationships between components of the application under design, and it may be harder for non-QCs to make such revisions.

\noindent\begin{center}
  \begin{tcolorbox}[colback=black!5, colframe=black!20, width=1.0\linewidth, arc=1mm, auto outer arc, boxrule=1.5pt]                       
             {{\textbf{Key Finding of RQ1:} The revision of ARQs is not prevalent on SO (31\%), and an ARQ revision starts soon after this question is posted (i.e., from 1 minute onward). Moreover, the revision of an ARQ occurs before and after this question receives its first answer/architecture solution, with most revisions beginning before the first architecture solution is posted. We found that both Question Creators (QCs) and non-QCs actively participate in ARQ revisions, with most revisions being made by QCs.}}
   \end{tcolorbox}
 \end{center}
\color{black}

\subsection{Missing and Further Provided Information in ARQs (RQ2)}\label{ResultsOfRQ2}
To answer RQ2, we used open coding \& constant comparison techniques (as described in Section \ref{dataAnalyis}) to investigate the missing and further provided information in ARQs. Our data analysis yielded 14 categories of missing and further provided information in ARQs. %Note that the sum of the counts of ARQs in Table \ref{CategoriesOfMissingAndFurtherProvidedInfo} is greater than the total number of the selected ARPs (\textcolor{black}{4,114}) because one post can miss several pieces of information. 
From Table \ref{CategoriesOfMissingAndFurtherProvidedInfo}, we found that \textit{design context} is the most missing and further provided type of information in ARQs when SO users describe their architecture problems, followed by \textit{component dependency}. \textit{Architecture concern} is the third category of most frequently missing and further provided information in ARQs. Moreover, \textit{architecture tactic}, \textit{architecture decision}, and \textit{software protocol} are the least discussed types of missing and further provided information in ARQs. The identified categories of RQ2 are presented below, and we provide an ARQ example for each detailed category.

\textbf{Design context} refers to the knowledge about the environments in which systems are expected to operate \citep{bedjeti2017modeling}. Design contexts are indispensable ingredients that can drive the architecture design of a system \citep{bedjeti2017modeling}, and while developers recognise them, they are not stated explicitly and are often tacit. As shown in Table \ref{CategoriesOfMissingAndFurtherProvidedInfo}, \textit{design context} is the most common category of missing and further provided information in ARQs. We found several types of design context that were missing and further provided in ARQs after being posted, such as application domain contexts (e.g., real-time, game, E-commerce, and bank systems) and platform contexts (e.g., Linux, Android, Windows, and iOS-based systems). 

\noindent\begin{center}
  \begin{tcolorbox}[colback=white!5, colframe=black!20, width=1.0\linewidth, arc=1mm, auto outer arc, boxrule=1.5pt] 
             {{
             \textbf{Commenter}: \color{black!70!white}\faCommenting\color{black}\hspace{0.5mm} Unfortunately, I fear that the answer depends heavily on the \textit{operating system} you are running on (...). \newline
             \textbf{Asker}: \color{black!70!white}\faCommenting\color{black}\hspace{0.5mm} Sorry, forgot to write it. It is \textit{Linux} (I've edited the question). (SO post \#7847089) 
             }}    
   \end{tcolorbox}
 \end{center}

\textbf{Component dependency}. A software system may be designed with several heterogeneous components (also known as modules), such as software components and hardware components components, which need to interact and communicate with each other to satisfy various architecture design concerns \citep{SA2012}. We collected several ARQs wherein askers failed to detail component or module dependencies in their architecture problem descriptions. 

\noindent\begin{center}
  \begin{tcolorbox}[colback=white!5, colframe=black!20, width=1.0\linewidth, arc=1mm, auto outer arc, boxrule=1.5pt]
             {{
             \textbf{Commenter}: \color{black!70!white}\faCommenting\color{black}\hspace{0.5mm} ‘The problem was, of course, that not all classes (or components) are used throughout the application equally’ OK, you didn't mention that :) (...). 
             
             \textbf{Asker}: \color{black!70!white}\faCommenting\color{black}\hspace{0.5mm} Edit: We have a layered architecture and we manage dependencies between the modules/layers by setting references between the Visual Studio projects/assemblies. To clarify, not only does Consumer B depend on B, but also on A, since B extends A. However, if A is defined in a different assembly than the assembly of B, then we also must add a reference to the assembly of A (the same applies to providers). But there is no actual need to add this reference, since the dependency on A follows from the fact that we depend on B and B extends A (...) (SO post \#18466069)
             }}    
   \end{tcolorbox}
 \end{center}

\textbf{Architecture concern} refers to high-level qualities (e.g., performance, availability, scalability, security) that a software system should address. We observed that askers often failed to clearly elaborate on the architecture concerns that their systems need to address in architecture problem descriptions. 

\noindent\begin{center}
  \begin{tcolorbox}[colback=white!5, colframe=black!20, width=1.0\linewidth, arc=1mm, auto outer arc, boxrule=1.5pt] 
             {{
             \textbf{Commenter}: \color{black!70!white}\faCommenting\color{black}\hspace{0.5mm} Are you sure you're not prematurely optimizing this? How long does the query take currently? I would guess it's less than 5 seconds if you have the right indexes and don't have millions of users. Running it once a minute may not be so bad.
             %What's the question or \textit{concern}? “Is this a good thing?” You will probably not get any good answers to such an open-ended question. 

             \textbf{Asker}: \color{black!70!white}\faCommenting\color{black}\hspace{0.5mm} The thing is we don't need to have millions of users because each user can have multiple data objects with their own specific time just like multiple alarms in a single phone for which the user needs to be notified. Also I am not doing this for my personal project but for a rather big company so we do have a large user base and also this needs to be \textit{scalable} (...). (SO post \#76595792) 
             %Sorry, I should have been more clear (edited now) (...) I'm just \textit{concerned} as to how this platform will \textit{scale} and how \textit{secure} it will be. Thanks. (SO post \#3993121) 
             }}    
   \end{tcolorbox}
 \end{center}

\textbf{Architecture patterns} (such as Model View Controller (MVC), Client-Server, and Publish-Subscribe patterns) are reusable solutions to commonly occurring problems in architecture design within given contexts \cite{SA2012}. Architectural patterns determine the overall structure and behavior of a software system \citep{buschmann1996pattern} and are typically selected early during development for achieving multiple system requirements (e.g., quality attributes) \cite{SA2012}. With our dataset, we found ARQs in which askers failed to state architecture patterns (in the case that their systems are being designed following certain architecture patterns) in architecture problem descriptions.  

\noindent\begin{center}
  \begin{tcolorbox}[colback=white!5, colframe=black!20, width=1.0\linewidth, arc=1mm, auto outer arc, boxrule=1.5pt] 
             {{
             \textbf{Commenter}: \color{black!70!white}\faCommenting\color{black}\hspace{0.5mm} That's a ``hook'' title if I've ever seen one (...). 
             
             \textbf{Asker}: \color{black!70!white}\faCommenting\color{black}\hspace{0.5mm} I have edited my question (...). Please note I am using \textit{MVVM} architecture. (SO post \#4552281)
             }}    
   \end{tcolorbox}
 \end{center}

 \textbf{Framework} (such as Django, Spring, Angular, WCF) is a partial architecture design (accompanied by code) that provides common services in particular domains~\citep{SA2012}. Moreover, frameworks consist of collections of reusable software components put together to ease development. We gathered several ARQs in which askers failed to clarify frameworks in architecture problem descriptions (in the case that the systems under design are following certain frameworks). 

 \noindent\begin{center}
  \begin{tcolorbox}[colback=white!5, colframe=black!20, width=1.0\linewidth, arc=1mm, auto outer arc, boxrule=1.5pt] 
             {{
            \textbf{Commenter}: \color{black!70!white}\faCommenting\color{black}\hspace{0.5mm} Can you provide more information (...) There are many factors to consider before jumping into a data mode (...). 
            
            \textbf{Asker}: \color{black!70!white}\faCommenting\color{black}\hspace{0.5mm} Edit for Suirtimed: (...) This is an SOA style project, using \textit{WCF} (...). (SO post \#4554925)
             }}    
   \end{tcolorbox}
 \end{center}

 \textbf{Requirements} refer to functionalities or features that a system should possess \citep{hull2005requirements}. The requirements need to be analyzed and mapped to their corresponding architectural elements (e.g., architectural components) during software architecture design \citep{hull2005requirements, casamayor2012functional}. However, we accumulated several scenarios in ARQs where askers failed to clarify the features/requirements of the systems or failed to relate/map these requirements to their corresponding architectural components.  

\noindent\begin{center}
  \begin{tcolorbox}[colback=white!5, colframe=black!20, width=1.0\linewidth, arc=1mm, auto outer arc, boxrule=1.5pt] 
             {{
             \textbf{Commenter}: \color{black!70!white}\faCommenting\color{black}\hspace{0.5mm} What are your \textit{requirements}? (...). 
             
             \textbf{Asker}: \color{black!70!white}\faCommenting\color{black}\hspace{0.5mm} Edit in question: My \textit{requirements} now are that I need a versatile architecture behind my core. \textit{Core should be there only to make everything work together, not do stuff like content managing or anything}, that's what modules are meant for (...).  I've imagined, that \textit{core would only take user requests. According to request, send them to the appropriate module, but the module again can request a library or something, that should go through the core. Everything has to work with no extra use. If the module doesn't need a database, then there shouldn't be a loaded library for the database}.(SO post \#3787028)
             }}    
   \end{tcolorbox}
 \end{center}

 \textbf{Types of component}. A system may be built using different types of components, such as User Interface (UI), Data Access Layer (DAL), and Business Logic Layer (BLL) components. This category collects ARQs in which askers failed to clarify types of components (that will constitute the system under design) and their descriptions in architecture problems. 

\noindent\begin{center}
  \begin{tcolorbox}[colback=white!5, colframe=black!20, width=1.0\linewidth, arc=1mm, auto outer arc, boxrule=1.5pt] 
             {{
             \textbf{Commenter}: \color{black!70!white}\faCommenting\color{black}\hspace{0.5mm}  (...) Can you update the question with some details? What kind of components (or applications) does it contain?
             
             \textbf{Asker}: \color{black!70!white}\faCommenting\color{black}\hspace{0.5mm} I've been asked for some clarification on the current project architecture. Without going into too much detail, think of a .NET web forms application that plugs into a \textit{layer of business logic} and integrates with multiple \textit{third-party vendor}. (SO post \#11118030) %These projects have several identical \textit{Controllers/Views} and \textit{_Layout} (MasterPage). (SO post \#14729684)
             }}    
   \end{tcolorbox}
 \end{center}

 \textbf{Programming language} includes ARQs in which askers failed to clarify programming language used or planned to be used in architecture implementation in architecture problem description. 

\noindent\begin{center}
  \begin{tcolorbox}[colback=white!5, colframe=black!20, width=1.0\linewidth, arc=1mm, auto outer arc, boxrule=1.5pt] 
             {{
             \textbf{Commenter}: \color{black!70!white}\faCommenting\color{black}\hspace{0.5mm}  We are going to need more information (...) What language will the site be developed in? 
             
             \textbf{Asker}: \color{black!70!white}\faCommenting\color{black}\hspace{0.5mm} Edit in question: The site is \textit{php} (...). (SO post \#4824667)
             }}    
   \end{tcolorbox}
 \end{center}

\textbf{Architecture diagram}: ARQs in this category missed to provide architecture diagrams (such as component diagrams) in architecture problem. 

 \noindent\begin{center}
  \begin{tcolorbox}[colback=white!5, colframe=black!20, width=1.0\linewidth, arc=1mm, auto outer arc, boxrule=1.5pt] 
             {{
             \textbf{Commenter}: \color{black!70!white}\faCommenting\color{black}\hspace{0.5mm}  If you use create C\# assemblies then you wouldn't be using interop to use them. Why don't you just continue the use of your Store Procedures? 
             
             \textbf{Asker}: \color{black!70!white}\faCommenting\color{black}\hspace{0.5mm} I assume that I'd have to create the COM wrappers as I couldn't access .NET managed dlls otherwise. See my edit for a \textit{diagram} of what I'm talking about. (SO post \#11582100) 
             }}    
   \end{tcolorbox}
 \end{center}

 \textbf{Database system}: Answerers would like askers to describe database systems in their ARQs so that these answerers can provide relevant or useful architectural solutions to those questions. However, askers sometime failed to provide such information. 

\noindent\begin{center}
  \begin{tcolorbox}[colback=white!5, colframe=black!20, width=1.0\linewidth, arc=1mm, auto outer arc, boxrule=1.5pt] 
             {{
             \textbf{Commenter}: \color{black!70!white}\faCommenting\color{black}\hspace{0.5mm}  What product (database) are you talking about? Oracle, SQL Server? (...) 
             
             \textbf{Asker}: \color{black!70!white}\faCommenting\color{black}\hspace{0.5mm} In response to a comment on this question, we are using \textit{SQL Server 2005} (...). (SO post \#916295)
             }}    
   \end{tcolorbox}
 \end{center}

\textbf{API}: An application programming interface (API), such as (such as REST or SOAP API) offer a way for two or more computer programs or components to communicate with each other. This category gathers ARQs in which APIs  were missed in architecture problem description. 

\noindent\begin{center}
  \begin{tcolorbox}[colback=white!5, colframe=black!20, width=1.0\linewidth, arc=1mm, auto outer arc, boxrule=1.5pt] 
             {{
             \textbf{Commenter}: \color{black!70!white}\faCommenting\color{black}\hspace{0.5mm} Do you mean the client-side web page will directly query your database, rather than querying a web-service in order to get data? 
             
             \textbf{Asker}: \color{black!70!white}\faCommenting\color{black}\hspace{0.5mm}  I am planning a website (Internet facing) and would like to generate the HTML directly from a Database using Database Web Server (with proxy). For, example, creating a \textit{SOAP web service} in anywhere database. (SO post \#5434190)
             }}    
   \end{tcolorbox}
 \end{center}

\textbf{Architecture tactics} (such as heartbeat, resource pooling) as design primitives provide a “\textit{means of satisfying quality-response measure by manipulating some aspect of a quality model trough architectural design decisions}”~\cite{bass2000quality}. Unlike architectural patterns that are related to multiple quality attributes, architectural tactics are used to address one specific quality attribute \cite{harrison2010architecture}. For example, architectural tactics for addressing performance, such as resource pooling, help optimize response time. The ARQs in this category missed to state architecture tactics in the architecture problem descriptions (in the case that the systems under design are already using certain architecture tactics).

 \noindent\begin{center}
  \begin{tcolorbox}[colback=white!5, colframe=black!20, width=1.0\linewidth, arc=1mm, auto outer arc, boxrule=1.5pt] 
             {{
             \textbf{Commenter}: \color{black!70!white}\faCommenting\color{black}\hspace{0.5mm} What about a shell script on the central node that pings them regularly?
             
             \textbf{Asker}: \color{black!70!white}\faCommenting\color{black}\hspace{0.5mm} Sorry, I should have mentioned that we may later need to send some performance metrics with the messages. Also, a ping will go out over TCP which may not be ideal. Edit in question: We may later need to also pass some text with the “\textit{heartbeat}” messages, so we want to develop the system keeping that in mind. (SO post \#5297178) %My main intent is to demonstrate the architecture and build process of an app with \textit{OAuth-based authentication} without disseminating my private keys all over the internet. Therefore, a need for a public code repository and a publicly visible build server. (SO post \#14585228)
             }}    
   \end{tcolorbox}
 \end{center}

\textbf{Architecture decision} is a description of a set of architectural additions, subtractions, and modifications to the architecture, the rationale, design rules, design constraints, and additional requirements that (partially) realize one or more requirements on a given architecture \citep{jansen2005SoftArch}. Architectural decisions play a crucial role in the design, implementation, evolution, reuse, and integration of architectures. This category gathers ARQs in which askers failed to state architecture decisions in architecture problem descriptions (in the case that the systems under design adhere or stick to already made architecture decisions).

 \noindent\begin{center}
  \begin{tcolorbox}[colback=white!5, colframe=black!20, width=1.0\linewidth, arc=1mm, auto outer arc, boxrule=1.5pt] 
             {{
            % \textbf{Commenter}: \color{black!70!white}\faCommenting\color{black}\hspace{0.5mm} The design issue is now clear since you added these decisions in your design. Thanks :) 
             
             \textbf{Asker}: \color{black!70!white}\faCommenting\color{black}\hspace{0.5mm} Edit in question: \textit{Decisions I've taken}: Use domain model for MVVM's model (removed PersonModel) Use external mappings for the same model to the database (removed PersonEntity added PersonMappings) (...). (SO post \#38431091) %Several people requested that I update this question with the results from our meeting. We debated back and forth the \textit{pros and cons} of doing this (using a single master database for all new applications) (...). We \textit{decided} to use some time with Microsoft to get their thoughts and platform-specific advice.  (SO post \#916295) 
             }}    
   \end{tcolorbox}
 \end{center}

 \textbf{Software protocols} %(such as TCP and UDP protocols) refer to a set of rule for systems or components of systems to communicate with each others. 
 This category of missing information collects ARQs in which asked failed to clarify software protocols (such as HTTP, TCP, UDP protocols, OAuth) used in architecture design in architecture problem description. 

\noindent\begin{center}
  \begin{tcolorbox}[colback=white!5, colframe=black!20, width=1.0\linewidth, arc=1mm, auto outer arc, boxrule=1.5pt] 
             {{
             \textbf{Commenter}: \color{black!70!white}\faCommenting\color{black}\hspace{0.5mm} Are you using \textit{tcp} or \textit{udp} connection?
             
             \textbf{Asker}: \color{black!70!white}\faCommenting\color{black}\hspace{0.5mm}  I think it was using tcp (...). (SO post \#32929731)
             }}    
   \end{tcolorbox}
 \end{center}
\color{black}

\small
\begin{longtable}{p{10em}p{22.3em}p{3em}}
\caption{Categories of missing and further provided information in ARQs after being posted} \label{CategoriesOfMissingAndFurtherProvidedInfo} \\\hline
\textbf{Category} & \textbf{Description} & \textbf{Counts} \\\hline   
\endfirsthead
{{\bfseries }}\\
\endhead
\endfoot
\hline
\endlastfoot
{Design context}                                  
                                               & Design context description, such as application domain context (e.g., game or E-commerce system) \cite{bedjeti2017modeling}) is missing in the architecture problem.
                                               & 860 \\\cline{1-3}                             
{Component dependency}
                                               & Detailed component or module dependency description is missing in architecture problem.
                                               & 605  \\\cline{1-3}                      
{Architecture concern}
                                               & Architecture concern description is missing in architecture problem. 
                                               & 516 \\\cline{1-3}  

{Architecture pattern}    
                                                & Architecture pattern (e.g., MVVM, MVC, Microservice) is missing in architecture problem description.
                                                & 462  \\\cline{1-3}          
{Framework} 
                                                & In use framework (e.g., WCF, ASP.NET, and Django) is missing in architecture problem description.
                                                & 425  \\\cline{1-3}

{Type of component} 
                                               & Type of component/module or layer (such as User Interface (UI), Data Access Layer (DAL), and Business Logic Layer (BLL)) is missing in architecture problem description.
                                               & 263  \\\cline{1-3}                                                                                              
{Requirement}
                                                & Functionalities or features that a system should possess are missing in architecture problem.
                                                & 213  \\\cline{1-3}                      

{Programming language}
                                                & The programming language used in architecture implementation is missing in architecture problem description. 
                                                & 181 \\\cline{1-3}                                               
{Architecture diagram}
                                                & Architecture diagram (e.g., component diagram) is missing in architecture problem description. 
                                                & 134  \\\cline{1-3}  
{Database system}
                                                & In use database system is missing in architecture problem description. %\textbf{Commenter}: What product (database) are you talking about? Oracle, SQL Server? (...)  \textbf{Asker}: In response to a comment on this question, we are using \textit{SQL Server 2005} (...). (SO post \#916295)
                                                & 122 \\\cline{1-3} 

{API}
                                                & In use API (e.g., REST or SOAP API) is missing in architecture problem description. 
                                                & 103    \\\cline{1-3}                                                                                                
{Architecture tactic} 
                                                & The architecture tactics (e.g., heartbeat) used in architecture is missing in architecture problem description. 
                                                & 87 \\\cline{1-3}                                                
{Architecture decision} 
                                                & Made architecture decision is missing in architecture problem description.  
                                                & 57 \\\cline{1-3}                                               
{Software protocol}
                                                & In use software protocol (e.g., OAuth, SSL, TCP) is missing in architecture problem description. %\textbf{Commenter}: \textbf{Asker}: Are you using \textit{tcp} or \textit{udp} connection?  \textbf{Asker}: I think it was using tcp (...). (SO post \#32929731)
                                                & 49 \\\cline{1-3}   
                                                                        
\end{longtable}
\normalsize

\noindent\begin{center}
  \begin{tcolorbox}[colback=black!5, colframe=black!20, width=1.0\linewidth, arc=1mm, auto outer arc, boxrule=1.5pt]            
             {{\textbf{Key Finding of RQ2:} A broad range (14 categories) of information is missing and further provided in ARQs after being posted, among which \textit{design context}, \textit{component dependency}, and \textit{architecture concern} are the top three categories of most frequently missing and further provided information in ARQs.
             }}    
   \end{tcolorbox}
 \end{center}
\color{black}

\subsection{Purposes of the Further Provided Information in ARQs (RQ3)}\label{ResultsOfRQ3}
Similar to RQ1, we used open coding \& constant comparison techniques described in Section \ref{dataAnalyis} to answer RQ3. Our data analysis yielded five categories of purposes for the further information provided in ARQs. Table \ref{CategoriesOfPurposesOfTheProvidedInfo} shows the five categories with their count information. More specifically, we found that \textit{clarify the understanding of architecture under design} is the most frequent purpose of the further provided information in ARQs, followed by \textit{improve the readability of architecture problem}. The identified categories of RQ3 are presented below, and we provide an ARQ example for each category.

\textbf{Clarify the understanding of architecture under design} refers to improving architecture problem description by adding certain important architectural element information, such architecture patterns or tactics, component dependencies, quality concerns, architecture diagrams to make the architecture under design more clear and understandable. 

\noindent\begin{center}
  \begin{tcolorbox}[colback=white!5, colframe=black!20, width=1.0\linewidth, arc=1mm, auto outer arc, boxrule=1.5pt] 
             {{
              \textbf{Commenter}: \color{black!70!white}\faCommenting\color{black}\hspace{0.5mm} Can you describe the dynamic behavior of illustrated components in domain-specific terms? E.g. “User X queries profile data and needs to query both services A and B”. Because currently, it is \textit{not very clear what those b2b requests are about}. 
              
              \textbf{Asker}: \color{black!70!white}\faCommenting\color{black}\hspace{0.5mm} I updated the question with \textit{more details regarding my current architecture} and what I'm trying to archive. Please take a look at the flows sections. (SO post \#41640621) %My application's architecture follows Project Silk's very closely. The data tier holds the repositories and model POCOs. The Business Logic layer holds the services. Inside these services, we map objects from the Data Tier to the Model objects in the business layer. (SO post \#12043856). %\textcolor{black}{\textit{\textbf{Commenter}: Please rephrase in terms of an actual question with more information about what the problem is (...). \textbf{Commenter}: I've done the edits (...)}. (SO post \#532399)}
             }}
   \end{tcolorbox}
 \end{center}

\textbf{Improve the readability of architecture problem} refers to improving architecture problem description by fixing grammar/typo issues or correcting text formatting issues (such as formatting the text for a better presentation) in order to make architecture problem more readable. 

\noindent\begin{center}
  \begin{tcolorbox}[colback=white!5, colframe=black!20, width=1.0\linewidth, arc=1mm, auto outer arc, boxrule=1.5pt] 
             {{
              \textbf{Commenter}: \color{black!70!white}\faCommenting\color{black}\hspace{0.5mm} I think you want to revise your last sentence, as both MVC and WebForms are ASP.NET (...). I hope you don't mind, \textit{I cleaned up some typos and stuff} to make the question more readable.

              \textbf{Asker}: \color{black!70!white}\faCommenting\color{black}\hspace{0.5mm} oops okay now it should be better. (SO post \#1035642)
             }}
   \end{tcolorbox}
 \end{center}

\textbf{Clarify the attempted architecture solution} refers to improving architecture problem description by stating the architecture solutions that have been tried to solve the problem.

\noindent\begin{center}
  \begin{tcolorbox}[colback=white!5, colframe=black!20, width=1.0\linewidth, arc=1mm, auto outer arc, boxrule=1.5pt] 
             {{
              \textbf{Commenter}: \color{black!70!white}\faCommenting\color{black}\hspace{0.5mm} \textit{What have you tried so far (solution)?} (...) Maybe solving a simple version of this problem will give you/us insight on how to solve your problem.
              
              \textbf{Asker}: \color{black!70!white}\faCommenting\color{black}\hspace{0.5mm} \textit{I add some description of what I have tried so far} at the bottom of the question. (SO post \#5221556) %I updated my possible solution. Of course the above solution is just for shared UUID/GUID across microservices. For full data, an event bus is used to distribute events and data across microservices in an asynchronous way (Event sourcing pattern). The main goal is always to avoid data duplication. Denormalization is archived with bounded context (This is critical of course). (SO post \#41640621). %I tried this out in non-razor land to see if the same is true, which it is. I had to use a template param in the View to get it to work. Here is that non-razor view (SO post \#15646812)
             }}
   \end{tcolorbox}
 \end{center}

\textbf{Clarify the intended use of the system under design} refers to improving architecture problem description by providing the intended use of an application under design.

\noindent\begin{center}
  \begin{tcolorbox}[colback=white!5, colframe=black!20, width=1.0\linewidth, arc=1mm, auto outer arc, boxrule=1.5pt] 
             {{
               \textbf{Commenter}: \color{black!70!white}\faCommenting\color{black}\hspace{0.5mm} Can you offer some more details on \textit{what the web application is used for} (...) everything depends on purpose (...).
               
               \textbf{Asker}: \color{black!70!white}\faCommenting\color{black}\hspace{0.5mm} I have edited and added some more details to it. \textit{It (web application) is for managing documents} (...). (SO post \#2332717)
             }}
   \end{tcolorbox}
 \end{center}

\begin{table*}[h]
\centering
\small
\footnotesize
\caption{Categories of the purposes of the further provided information in ARQs}
\label{CategoriesOfPurposesOfTheProvidedInfo}
\begin{tabular}{m{7cm}m{1cm}<{\centering}} 
\hline
\textbf{Category}                                              &\textbf{Counts}    \\ \hline

Clarify the understanding of architecture under design         & 2767 \\\cline{1-2} 

Improve the readability of architecture problem                & 797  \\\cline{1-2}   

Clarify the attempted architecture solution                    & 414 \\\cline{1-2}

Clarify the intended use of the system under design            & 136 \\\cline{1-2} 

\textcolor{black}{Compliance with organizational policy}        & \textcolor{black}{1} \\\cline{1-2} 

\end{tabular}
\end{table*}

\normalsize
\noindent\begin{center}
  \begin{tcolorbox}[colback=black!5, colframe=black!20, width=1.0\linewidth, arc=1mm, auto outer arc, boxrule=1.5pt]            
             {{\textbf{Key Finding of RQ3:} \textit{Clarify the understanding of architecture under design} is the most frequent category of purposes of the further provided information in ARQs.
             }}  
   \end{tcolorbox}
 \end{center}\color{black}

\subsection{Impact of the Further Provided Information in ARQs on the Answers (RQ4)}\label{ResultsOfRQ4}
Analogous to the previous RQs, we used open coding \& constant comparison to answer RQ4 (as described in Section \ref{dataAnalyis}). The outputs of our data analysis generated four categories of the impact of the further provided information in ARQs on the answers. We provide these four categories with their count information in Table \ref{CategoriesOfImpactsOfTheFurtherProvidedInfo}, which shows that \textit{make an architecture solution useful} and \textit{make an architecture solution informative} are the top two most frequent impacts of the further provided information in ARQs on the answers. Moreover, the identified categories of RQ4 are presented below, and we provide an ARQ example for each category.

\textbf{Make an architecture solution useful} denotes a solution that provides factual and valuable information or knowledge that could be turned into practice during development. It goes beyond the mere provision of theoretical information or personal interpretations to encompass the effectiveness of that information or knowledge in practice \citep{hu2022correlating}.

 \noindent\begin{center}
  \begin{tcolorbox}[colback=white!5, colframe=black!20, width=1.0\linewidth, arc=1mm, auto outer arc, boxrule=1.5pt] 
             {{
             \textbf{Edit in the question}: \color{black!70!white}\faQuestionCircle\color{black}\hspace{0.5mm} More information about my app: My application is a management software and manage user registration, products, order and other things like those. In practice it contains a lot of read entity->edit->save entity or create->edit->save operations (...). 

             \textbf{Part of answer after edit(s) in the question}: \color{black!70!white}\faThumbsUp\color{black}\hspace{0.5mm} If your service layer duplicates DAO, you are not using service layer at all. I made the same mistake in few of my applications (...). The service layer should be interface for your application (...). Your swing app/web client will then be a client to your core component, you can then switch them without any limitations (because everything that modifies data is in core). Even in simple CRUD application, some kind of actions should be in service layer, for example, validation and authorization (...). %The other thing that you should/can do on service layer is authorization (is the user in this role permitted to delete this entry). Then you will have to have an service layer for almost all of your entities, because simple user should be able to edit (delete) his entries, but probably should not delete other users. But user in role admin can do this (...).
             
             \textbf{Comment under the answer}: \color{black!70!white}\faCommenting\color{black}\hspace{0.5mm} This answer is \textit{very practical and useful}. Thanks a lot! (...). (SO post \#4270461)
             }}    
   \end{tcolorbox}
 \end{center}

 \textbf{Make an architecture solution informative} denotes a solution that provides users (e.g., askers) with the information or knowledge to enhance their understanding of software architecture concepts. Contrary to the previous category, the solutions in this category are not necessarily meant to be readily useful in practice.  

\noindent\begin{center}
  \begin{tcolorbox}[colback=white!5, colframe=black!20, width=1.0\linewidth, arc=1mm, auto outer arc, boxrule=1.5pt] 
             {{
             \textbf{Edit in the question}: \color{black!70!white}\faQuestionCircle\color{black}\hspace{0.5mm} Client: The client can run on the same system as the server, manages all the IO, manages business logic for actual task execution (bad idea?), and reports status through Remoting/WCF via notification events. 

             \textbf{Part of answer after edit(s) in the question}: \color{black!70!white}\faThumbsUp\color{black}\hspace{0.5mm} (...) Generally it is a bad idea to have a single point of failure in a system (one special PC is the ``server'' and all others need to communicate to it). But in your case it could very well be that it does not matter. Read up on distributed systems design. The API you choose is not the most important thing, I'd say. First you need to plan the architecture, and the behavior you want to get in various scenarios (like one of the PCs goes AWOL). Just a quick reaction from reading your requirements, I would look into a distributed pub/sub system. How you implement the pub/sub is up to you. You could use MSMQ, or WCF and WS-Eventing (...).
             
             \textbf{Comment under the answer}: \color{black!70!white}\faCommenting\color{black}\hspace{0.5mm} 
             I learned a ton from your answer, that is definitely \textit{informative} input. (SO post \#683750) 
             %\textit{This was informative response, thank you. It very well is the case here that if one goes down it's not a big deal (...)}. (SO post \#683750)%(SO post \#28021807} 
             }}
   \end{tcolorbox}
 \end{center}

  \textbf{Make an architecture solution relevant} denotes a solution that aligns with the ARQ rather than a solution with irrelevant content or information. According to the previous study by \cite{xu2017answerbot}, if a solution contains software-specific entities or concepts (such as architecture patterns, frameworks, and APIs) mentioned in the question, the solution is likely relevant to the question. Contrary to the previous categories, the solutions in this category are not necessarily meant to be readily useful in practice or informative. 
  
\noindent\begin{center}
  \begin{tcolorbox}[colback=white!5, colframe=black!20, width=1.0\linewidth, arc=1mm, auto outer arc, boxrule=1.5pt] 
             {{
             \textbf{Edit in the question}: \color{black!70!white}\faQuestionCircle\color{black}\hspace{0.5mm} I feel that the reason for this is that the data is pretty huge. I am not aware of how Big Data applications are managed to deliver high performance. Please advise. Problems: the tool/application is very badly slow when it comes to aggregations, finds, etc. 
             
             \textbf{Part of answer after edit(s) in the question}: \color{black!70!white}\faThumbsUp\color{black}\hspace{0.5mm} What I would do is a detailed monitoring, at least the following: Database \& DB Process Level (my first guess, that this is the critical part): Monitor memory usage and CPU load for your Mango DB process. Have look at the caching strategy. Node.js application: If there are multiple instances of the node.js app, track this for all instances. If you see that you have a high load on this app that there is a need to act here (increasing instances, splitting it into separate apps (e.g. import as a separate app)). 
               
             \textbf{Comment under the answer}: \color{black!70!white}\faCommenting\color{black}\hspace{0.5mm} Thanks for taking the time to provide such a detailed and incredibly \textit{relevant} answer (SO post \#69200534) 
             }}
   \end{tcolorbox}
 \end{center}

\textbf{Make an architecture solution complete}: A user may ask more than one question (sub-question) in one single ARQ. This category of impacts refers to the solutions that address all sub-questions (e.g., architecture concerns) asked in the ARQ.

\noindent\begin{center}
  \begin{tcolorbox}[colback=white!5, colframe=black!20, width=1.0\linewidth, arc=1mm, auto outer arc, boxrule=1.5pt] 
             {{
             \textbf{Edit in the question}: \color{black!70!white}\faQuestionCircle\color{black}\hspace{0.5mm} I have a somewhat complicated situation. I'll try my best to explain both in pictures/diagrams and words. Here's the current system architecture (...) and the proposed architecture I envision is this: (...). 

             \textbf{Part of answer after edit(s) in the question}: \color{black!70!white}\faThumbsUp\color{black}\hspace{0.5mm} (...) The initial plan of using PyTables as an intermediary layer between your other elements and the data files seems solid. (...). Additionally, on your diagram you have components of different types all using the same symbols. It makes it a little bit difficult to analyze the expected performance and efficiency. 
             
             \textbf{Comment under the answer}: \color{black!70!white}\faCommenting\color{black}\hspace{0.5mm} +1 for the most \textit{complete}, comprehensive response I've ever seen. You are right of course, I should be thinking of a way to decouple the systems from each other (...). (SO post \#1953731)
             
             }}
   \end{tcolorbox}
 \end{center}

\begin{table}[h]
\centering
\small
\footnotesize
\caption{Categories of the impact of the further provided information in ARQs on the answers/architecture solutions}
\label{CategoriesOfImpactsOfTheFurtherProvidedInfo}
\begin{tabular}{m{7cm}m{0.9cm}<{\centering}}
\hline
\textbf{Category}                                                      &\textbf{Counts}     \\ \hline

Make an architecture solution useful                                   & 1912 \\\cline{1-2} 
                                                       
Make an architecture solution informative                              & 948 \\\cline{1-2} 

Make an architecture solution relevant                                 & 511 \\\cline{1-2} 

Make an architecture solution complete                                 & 181 \\\cline{1-2}

\textcolor{black}{Others (e.g., receiving more answers/architecture solutions)}           & \textcolor{black}{1} \\\cline{1-2} 
\end{tabular}
\end{table}

\normalsize
\noindent\begin{center}
  \begin{tcolorbox}[colback=black!5, colframe=black!20, width=1.0\linewidth, arc=1mm, auto outer arc, boxrule=1.5pt]                       
             {{\textbf{Key Finding of RQ4:} \textit{Make an architecture solution useful} and \textit{make an architecture solution informative} are the top two most frequent categories of impact of the further provided information in ARQs on the answers.}} 
   \end{tcolorbox}
\end{center}

\color{black}
\subsection{Practitioners' Perceptions of the Categorization of ARQ Revisions (RQ5)}\label{ResultsOfRQ5}

To evaluate the validity, accuracy, and quality of our categorization of ARQ revisions, we conducted semi-structured interviews with experienced software practitioners. The interview responses were analyzed using open coding and constant comparison, revealing strong support for the quality and usefulness of the identified categories. As described earlier, our evaluation was organized into four phases: I. \textit{Evaluation of Categories of Missing and Further Provided Information in ARQs}, II. \textit{Evaluation of Categories of the Purposes of the Further Provided Information in ARQs}, III. \textit{Evaluation of Categories of the Impact of the Further Provided Information in ARQs on the Answers}, and VI. \textit{Practical Relevance and Applicability of the Findings to Developers and Moderators}. Firstly, all recruited participants reported using SO and other Q\&A sites to support their software architecture tasks. When asked, “\textit{How frequently do you use Stack Overflow and other Q\&A sites for architecture-related tasks?}” (GQ5), \textcolor{black}{6 out of 11 participants (55\%) reported using the Q\&A sites often, while 5 participants (45\%) stated that they sometimes rely on these platforms. Overall, the findings suggest that Q\&A sites play a regular role in practitioners’ architectural decision-making processes.}
%Firstly, all recruited participants reported using SO and other Q\&A sites to support their software architecture tasks. When asked, “\textit{How frequently do you use Stack Overflow and other Q\&A sites to do architecture tasks?}” (GQ5), most participants indicated that they often or sometimes rely on these platforms. On average, participants acknowledged that these resources play a regular role in their architectural decision-making process. 

\subsubsection{Phase I: Categories of Missing and Further Provided Information in ARQs}

To assess the \textit{categories of missing and further provided information in ARQs}, we asked participants three interview questions (Q6–Q8) after presenting them with the 14 categories identified in our empirical study (see Table~\ref{CategoriesOfMissingAndFurtherProvidedInfo}) as commonly missing or subsequently added information after ARQs are posted. First (Q6), we asked whether these categories were meaningful, reliable, and helpful in understanding what is typically missing or later added in ARQs. All participants agreed, emphasizing that the categories effectively and reliably support the identification and interpretation of such information. For example, one participant (P9) stated:  
\faHandORight \hspace{0.5mm} ``\textit{These categories are meaningful, trustworthy, and reflective of the realities observed on Stack Overflow. Many posts are initially unclear, incomplete, or vague, and as a result, important information is often missing or only added through subsequent revisions. These categories facilitate the recognition of patterns in how users revise their posts, specifically, what information is often missing at first and how users attempt to address those gaps. Overall, they assist in understanding both the nature of missing content and the types of information that tend to be added post-submission}''. Next (Q7), we asked whether the categories comprehensively capture all types of missing and further provided information in ARQs. All participants affirmed the comprehensiveness of the categorization and expressed satisfaction with the current scheme. Regarding Q8, which asks whether any additional categories should be included, no new categories were suggested.

\subsubsection{Phase II: Categories Related to the Purpose of the Further Provided Information in ARQs}

Similar to the evaluation in Phase I, we explored the \textit{categories related to the purpose of the further provided information in ARQs} through three interview questions (Q9–Q11). Before asking these questions, we presented participants with the four categories identified in our empirical study (see Table~\ref{CategoriesOfPurposesOfTheProvidedInfo}) as commonly representing the purposes behind further information added to ARQs after posting. In Q9, we asked whether these categories were meaningful, reliable, and helpful in understanding the purposes of the further provided information. All participants agreed that the categorization reliably and effectively captured common motivations behind post edits. In Q10, we asked whether the categories comprehensively cover all types of purposes for the further information added to ARQs. \textcolor{black}{10 out of 11 participants (91\%)} affirmed the comprehensiveness of the current scheme and expressed satisfaction with it. However, when asked in Q11 whether any additional categories should be included, one participant (P5) proposed one additional purpose category “compliance with organizational policy” (e.g., when developers revise ARQs to anonymize proprietary details or rephrase content due to internal policy). While this was viewed as less frequent, we incorporated this new category into Table \ref{CategoriesOfPurposesOfTheProvidedInfo}. \faHandORight \hspace{0.5mm} ``\textit{At a high level, the current categories already capture the main purposes quite well, like enhancing clarity in the architecture being designed. However, one possible addition might be revisions made specifically to align the question with organizational or project-specific guidelines, like when developers tweak ARQs to anonymize proprietary details or rephrase content due to internal policy}''.

\subsubsection{Phase III: Categories of Impact of the Further Information Provided in ARQs on Answers}

To evaluate the \textit{categories of impact of the further information provided in ARQs on answers}, we asked participants four interview questions (Q12–Q15). Before this, we presented them with the four impact categories identified in our empirical study (see Table~\ref{CategoriesOfImpactsOfTheFurtherProvidedInfo}) as commonly reflecting how such information influences answers. In response to Q12, which asks whether the categories are meaningful, reliable, and helpful in understanding the impact of ARQ revisions on answers, all participants agreed. They emphasized that the categories effectively and reliably capture how revisions influence the quality and content of answers. When asked in Q13 whether the provided categories comprehensively capture all types of impact of the further information in ARQs after posting, \textcolor{black}{10 out of 11 participants (91\%)} affirmed their comprehensiveness. However, in Q14, when asked whether any additional impact categories could be included, one participants (P6) suggested a new category ``receiving more answers/architecture solutions''. For instance, participant (P6) remarked:  
\faHandORight \hspace{0.5mm} ``\textit{Maybe this can be added? 'receiving more answers/architecture solutions?'}''. We added this new category into Table \ref{CategoriesOfImpactsOfTheFurtherProvidedInfo}. Regarding Q15, which asked whether revisions in ARQs improve the quality of answers on SO (in terms of relevancy, practicality, completeness, and informativeness), all participants agreed that revisions, such as adding architectural details, clarifying the design context, or specifying system requirements, substantially enhance the quality of responses. \textcolor{black}{Moreover, 6 out of 11 participants (55\%)} noted that such edits often attract more expert engagement and foster higher-quality discourse. As one participant (P3) explained:  
\faHandORight \hspace{0.5mm} ``\textit{When users revise their architectural questions, often by clarifying the design context, concerns, or elaborating on system requirements, they provide answerers with a clearer understanding of the problem space. Based on experience, such revisions significantly improve the quality of answers in terms of relevancy, practicality, completeness, and informativeness (...)}''.

\subsubsection{Phase IV: Practical Relevance and Applicability of the Findings to Developers and Moderators}

Finally, when asked whether the findings assist practitioners (e.g., developers, Stack Overflow moderators) in improving the quality of ARQs (Q16), participants broadly agreed that the categorization framework proposed in this study can enhance question quality and support community moderation. They noted that the categories not only clarify why and how revisions are made but also serve as a practical guide for crafting clearer and more complete ARQs on platforms like SO. As one participant (P1) remarked: \faHandORight \hspace{0.5mm} ``\textit{As a software architect, my answer is `Yes'. These findings can greatly assist both practitioners and the Stack Overflow community in enhancing the quality of question posts, particularly those concerning software architecture. By understanding the categories that capture key design information in ARQs, developers or architects can take deliberate steps to improve the clarity and relevance of their questions (...)}''. %These findings affirm the value of the categorization scheme and suggest its applicability in supporting both practitioners and future tool development aimed at enhancing architectural knowledge sharing in online communities, such as SO.

\noindent\begin{center}
  \begin{tcolorbox}[colback=black!5, colframe=black!20, width=1.0\linewidth, arc=1mm, auto outer arc, boxrule=1.5pt] 
             {\textbf{Key Finding of RQ5:} 
             \textcolor{black}{The categorization of ARQ revisions was perceived by practitioners as meaningful, reliable, and comprehensive in capturing how ARQs evolve on SO. Participants confirmed that the categories effectively reflect common types, purposes, and impacts of ARQ revisions. While the existing categorization was largely validated, a few participants proposed additional categories, such as organizational constraints. Overall, these findings highlight the practical relevance of the proposed scheme in supporting architectural knowledge sharing and guiding ARQ revision practices. The results suggest that the categorization is not only valuable to practitioners but also applicable for informing future tool development aimed at enhancing architectural knowledge exchange in online developer communities.}
             } 
   \end{tcolorbox}
\end{center}

\color{black}

\section{Discussion}\label{Discussion}
For each RQ, we first interpret the key findings and then discuss their implications for researchers (indicated with the \faLeanpub \hspace{0.5mm} icon) and SO users (indicated with the \faMale \hspace{0.5mm} icon). 

\textbf{The prevalence of ARQ revisions (RQ1)}.
The results of RQ1 reveal that both QCs (Question Creators) and non-QCs actively participate in ARQ revisions, with most revisions being made by QCs. During our data analysis, we observed that non-QCs are more likely to help \textit{improve the readability of architecture problem}, such as fixing grammar/typo issues and text formatting (see Table \ref{CategoriesOfPurposesOfTheProvidedInfo}). On the other hand, QCs mostly perform significant ARQ revisions, such as providing important architectural element information (e.g., design contexts, architecture patterns, architecture decisions, and requirements) in order to \textit{clarify the understanding of architecture under design}. One possible reason for our abovementioned observation is that performing significant ARQ revisions (e.g., providing component dependencies) requires deep knowledge of these questions, and it may be harder for non-QCs to make such revisions. \textcolor{black}{Moreover, from the results of RQ1, we found that the revision of ARQs is not prevalent on SO (only 31\%), but when it does occur, it tends to happen very early, often within the first few hour after posting. This result suggests that early revisions may be a sign of QCs taking initiative to ensure that their ARQs are clear and well-structured. \faLeanpub \hspace{0.5mm} This is an important finding for the Software Engineering (SE) community, as it highlights the need to support early-stage ARQ refinement. For instance, \faLeanpub \hspace{0.5mm} the SE community could leverage this to promote more effective self-editing practices before askers seek external help. \faLeanpub \hspace{0.5mm} Additional, for platforms like SO, this finding can inform the development of moderation or user-support tools aimed at assisting users in revising their questions at the early stages, before community feedback is provided. \faLeanpub \hspace{0.5mm} Moreover, platform designers and tool builders could develop features that prompt or guide users to revise unclear or incomplete questions shortly after posting. \faLeanpub \hspace{0.5mm} Furthermore, automated tools could be trained to detect ARQs likely in need of revision based on early indicators (e.g., missing design context or unclear architecture under design) and suggest improvements proactively.} 

\textbf{Missing and further provided information in ARQs (RQ2)}.
As shown in Table \ref{CategoriesOfMissingAndFurtherProvidedInfo}, a broad range (14 categories) of information is often missed when SO users describe or ask ARQs. \textit{Design context} is the most frequently missing and further provided information in ARQs. One potential reason could be that most of the answerers do consider design context as one of the indispensable ingredients that can drive the architecture design of a system \citep{bedjeti2017modeling}, and the answerers would like the askers to reveal the design concerns along with the design contexts in their ARQs. Moreover, the answerers prefer the askers to provide a brief description of their project background so that the answerers can provide potential architecture solutions with pros and cons based on the given design concerns and design contexts \citep{bedjeti2017modeling}. \faLeanpub \hspace{0.5mm} Whilst we know that SO users discuss design contexts along with design concerns when asking and revising ARQs (see Table \ref{CategoriesOfMissingAndFurtherProvidedInfo}), there have been very few studies on mining design contexts in SO (e.g., \cite{bi2018architecture}) to support architecture design, which is interesting for the software architecture community to further explore. \textit{Component dependency} is the second most frequently missing and further provided information in ARQs. One reason could be that \textit{component dependency} may occur almost in all types of systems and SO users (answerers) always want to know the relationships between components/modules in the architecture problem description. Moreover, knowing the relationships between components is beneficial for the answerers to provide suitable and relevant answers to ARQs since when a component depends on one another, several factors such as downtime, deployment, performance, and security issues need to be considered specifically in a large and complex system. \faLeanpub \hspace{0.5mm} The software architecture community should work towards this area (e.g., approaches and tools for mining component dependencies from diverse sources, including Stack Overflow~\citep{karthik2019automatic}). Another type of frequently missing and further provided information in ARQs is \textit{architecture concern}, indicating that SO users (askers) often miss or forget to clarify architecture concerns when describing their architecture problems, and this may affect their ARQ answering time (e.g., taking longer to get answers) or remain unanswered. %\textit{Requirement} is the fourth most frequently missing and further provided information in ARQs. Analyzing requirements to relate and map them to their corresponding architectural elements (e.g., architectural components) has become one of the major challenges faced by architects \citep{souza2019deriving}. Thus, it is not surprising that \textit{requirement} comes in the top four most missing and further provided information in ARQs. 

This study is the first of its kind to investigate the revisions of information provided in architecture-related posts (i.e., ARQs) in Q\&A sites (i.e., SO), and \faLeanpub \hspace{0.5mm} further research is needed on the revision of architecture-related posts, such as tools for helping \faMale \hspace{0.5mm} SO users during the revisions of ARQs. %\textcolor{black}{\faLeanpub \hspace{0.5mm} the architecture research community could further investigate the reported categories of missing and further provided information in ARQs and provide challenges and difficulties users (e.g., developers) face when asking and revising ARQs on Q\&A sites and consequently, provide practical contributions.}
Also, \faMale \hspace{0.5mm} SO users should be more careful when asking ARQs since they tend to miss a wide range of information (see Table \ref{CategoriesOfMissingAndFurtherProvidedInfo}) in their ARQs. 

\textbf{Purposes of the further provided information in ARQs (RQ3)}.
As shown in Table \ref{CategoriesOfPurposesOfTheProvidedInfo}, the purposes of the further provided information in ARQs can be classified into four categories, in which \textit{clarify the understanding of architecture under design} is the most common purpose of the further provided information in ARQs. An architecture of a system should be understood before any activity (e.g., architecture synthesis) is performed. The information gained during architectural understanding can be used as input for other activities (e.g., architecture implementation, maintenance and evolution) \citep{li2013application}. Understanding an architecture under design in architecture problem descriptions can help SO users (answerers) to acquire thorough knowledge about the architecture, such as comprehending the elements of the architecture (e.g., architecture concerns and design contexts) and their relationships (e.g., component dependencies), so that they (answerers) can provide quality solutions (e.g., useful architecture solutions, see Table \ref{CategoriesOfImpactsOfTheFurtherProvidedInfo}) to the architecture problems. However, it is not easy for SO users to describe and clarify their architecture under design in the architecture problem description, as an asker replied to the this comment: “\textit{At the risk of sounding glib, you need to try to make this question shorter}”. \textbf{Asker}: “\textit{(...) Sorry for the long post, it's quite difficult to describe my problem}” (SO post \#829597). 

\faLeanpub \hspace{0.5mm} Researchers should pay more attention to the most frequent purpose (i.e., \textit{clarify the understanding of architecture under design}) of the further provided information in ARQs. Specifically, researchers might conduct a practitioner survey to further investigate the obstacles (e.g., new emerging architectural concepts, limited knowledge in the construction of ARQs) that prevent SO users from asking ARQs that are clear and understandable, and establish new approaches and guidelines that could improve the construction of ARQs. Moreover, similar to the finding by~\cite{zhu2022empirical}, we observed that information discussed in comments posted under ARQs plays a key role in the improvement of the quality of the information shared in SO. For instance, during our data analysis, we observed that the purposes of missing and further provided information in ARQs are often discussed in the comments. Previous studies have also observed similar phenomena in comments posted under the answers \citep{readingAnswers2019}. However, several existing studies mainly considered the information contained in SO posts, specifically, in the questions and answers, to retrieve architectural information. For example, a domain-specific automatic search approach has been proposed by \cite{soliman2018improving} to improve the search of architectural information in SO. \cite{bi2021mining} developed a semi-automatic approach to mine posts from SO and structured the design relationships between architectural tactics and quality attributes in practice to help software engineers better make design decisions. While these approaches mine the content of questions and answers to retrieve relevant information (e.g., architectural information) for development, they do not leverage the information that is contained in comments under the questions. \faLeanpub \hspace{0.5mm} Thus our study results suggest that empirical studies on exploring the information shared in comments and extracting informative comments (e.g., comments that discuss about the improvement of ARQs) are much needed. %By understanding the information discussed in comments posted under ARQs on SO, we can design mechanisms on SO to better support the community effort in improving the quality of information shared in its ARQs. 

\textbf{Impact of the further provided information in ARQs on the answers (RQ4)}.
As shown in Table \ref{CategoriesOfImpactsOfTheFurtherProvidedInfo}, there are four categories of the impact of the further provided information in ARQs on the answers, such as \textit{make an architecture solution useful}. During our data analysis, we observed that SO users (e.g., askers) revised ARQs for the sake of quality answers/architecture solutions, thus, all the identified categories (in Table~\ref{CategoriesOfImpactsOfTheFurtherProvidedInfo}) are related to the answer/solution quality \citep{zhu2009multi}. Depending on the goal of askers, the quality of the answer can be characterized by several factors. As shown in Table \ref{CategoriesOfImpactsOfTheFurtherProvidedInfo}, askers or other types of SO users (e.g., answerers) have different views on the quality of the answers when revising ARQs in SO. Thus, all the four categories of impact reported in Table \ref{CategoriesOfImpactsOfTheFurtherProvidedInfo}, such as \textit{make an architecture solution informative} and \textit{make an architecture solution relevant} are of importance, and \faMale \hspace{0.5mm} SO users should rely well on these impacts when asking and revising architecture-related posts in order to have quality architecture solutions. Therefore, we provide the following suggestions for SO users when asking and revising ARQs with the likelihood getting quality architecture solutions according to the study results. In additional, according to the study results, \faMale \hspace{0.5mm} we provide a structured template (see Table \ref{StructuredTemplateForPostingARQs}) with the key architectural elements to better support SO users posting ARQs. Note that the four architecture elements were chosen according to our experience in the architecture knowledge shared in SO.

\textit{Improve the understanding and readability of ARQs to receive useful, informative, complete, and practical architecture solutions}: \faMale \hspace{0.5mm} We suggest SO users to describe (when asking or revising) ARQs that are clear and easy to read in order to receive quality architecture solutions, including (1) useful architecture solutions (i.e., solutions that address the questions \citep{zhu2009multi}), (2) informative architecture solutions (i.e., solutions with suitable information in relation to the questions \cite{zhu2009multi}), (3) complete architecture solutions (i.e., solutions that completely answer the whole question \cite{zhu2009multi}), and (4) practical architecture solutions (i.e., solutions that provide facts, resources, personal experience, or real cases rather than ideas or imagination). This can be done, for example, by clearly providing an overall understanding of the architecture under design and specifically clarifying the design concerns in their ARQs, as well as detailed information on component dependencies together with interfaces. Also, \faMale \hspace{0.5mm} we recommend SO users provide information about other important elements of the architecture, such as architecture patterns, requirements, attempted architecture solutions, and design contexts, which are critical for potential SO answerers to correctly understand your architecture questions/problems. 

\begin{table}[h!]
%\small
\footnotesize
\caption{A structured template for posting ARQs}
\label{StructuredTemplateForPostingARQs}
\begin{tabular}{m{2cm}m{9.2cm}} 
\hline
\textbf{Key arch. element}                                     & \textbf{Description}     \\ \hline

Title of ARQ
                                                               & Write a title that summarizes the specific architecture problem (e.g., how can I extend my architecture for a Web application?). \\\cline{1-2}
Design context
                                                               & Provide the knowledge about the environments in which a system is expected to operate in (e.g., application and hardware contexts) \citep{bedjeti2017modeling}. \\\cline{1-2}          
Architecture concern 
                                                               & Describe the architecture concerns that your architecture needs to address (e.g., performance, reliability, and security).\\\cline{1-2}  
Architecture decision 
                                                               & Elaborate on already made architecture decisions with their rationale (e.g., component decisions). \\\cline{1-2} 
Attempted architecture solution 
                                                               & Elaborate on the potential architecture solution that you have tried and the design results accordingly. \\\cline{1-2} 
Architecture diagram 
                                                               & Include an architecture diagram (e.g., component diagram) which may help better understand the architecture problem.\\\cline{1-2}
                                                               
\end{tabular}
\end{table}

\section{Threats to Validity}\label{ThreatValidity}
We discuss the threats to the validity of our study by following the guidelines for empirical studies \citep{wohlin2012experimentation}, and how these threats were mitigated in our research.

\subsection{Construct validity} 

Construct validity concerns if the theoretical and conceptual constructs are correctly interpreted and measured \citep{wohlin2012experimentation}. One threat to the construct validity is concerned with the selection of search terms used to mine ARPs in SO. We used the search terms, i.e., “architect*”, to identify the related posts in SO (see Section \ref{Methodology}), and this is a threat to the construct validity in our study because relying solely on the keyword “architect*” may have excluded relevant posts that use alternative terminology, such as “design”. To reduce this threat, we first conducted a pilot search and observed that SO users use the term “design” mostly in the programming context (e.g., singleton design pattern\footnote{\url{https://tinyurl.com/vaak5a3c}}). In addition, as mentioned in Section \ref{Methodology}, using the search terms (e.g., “architecture”, “architecting”, and “architectural”) to only search exclusively through tags can be ineffective, because tags can be sometimes less informative \citep{chen2019modeling} (see the example in Section \ref{Methodology}). Thus, we decided to add the title and body of the questions to the search. In this way, we sought to minimize the risk of missing ARPs that use incorrect or irrelevant tags. Finally, we gathered 13,205 ARPs. To filter the 13,205 ARPs and get relevant ARPs for answering the RQs (i.e., RQ1, RQ2, RQ3, and RQ4), we defined a set of inclusion and exclusion criteria (see Table \ref{CriteriaForARPsWithRevisionInfo}) and followed a thorough post filtering process (as described in Section \ref{Methodology}), and finally got 4,114 ARPs that were used to answer the RQs. %Therefore, we believe that we have sufficiently minimized this threat.

\textcolor{black}{Although we attempted to mitigate this threat by performing a pilot search and applying the search to both the title and body of the questions, we still needed to quantify false positives identified when using the term ``design'', including specific numbers and percentages, to show the extent of the threat to the construct validity. Specifically, we wanted to perform an explicit analysis of the number of posts that may have been missed by limiting the search to ``architect*'' keywords. To achieve this, we downloaded and utilized the current SO data dump (i.e., Stack Exchange data dump on March 3, 2025\footnote{\url{https://archive.org/details/stackexchange_20250331}}). This data dump is a snapshot of the underlying database used by SO and it stores all the information for the questions, answers, tags, comments, votes, and user histories in XML files (e.g., \texttt{Posts.xml}). We used \texttt{Posts.xml} file, which stores the questions and answers of all the SO posts, as the basic to estimate how many ARPs we missed due to limiting the search to the ``architect*'' terms in our study. According to the SO data dump of March 3, 2025, there are 24 million questions (posts) and 36 million answers\footnote{\url{https://stackexchange.com/sites##oldest}}. We then used the power statistics and calculated a representative sample size of these 24 million posts. With a 95\% confidence level and 3\% margin of error, the representative sample size calculated is 1,068 posts. Afterwards, we randomly selected 1,068 posts from the 24 million posts and manually checked them for calculating how many ARPs we might have missed due to limiting the search to the “architect*” terms during the search of ARPs. Specifically, the first author labeled the 1,068 posts to determine which of the posts are ARPs. The second author checked and validated the labeling results. The disagreements were resolved in a meeting to improve the reliability of the labeling results. Based on our manual labeling, we found that out of the 1,068 posts, only 21 were ARPs (i.e., the true positives), wherein 14.3\% (i.e., 3 out of 21 ARPs) do not contain “architect*” terms and 85.7\% (i.e., 18 out of 21 APRs) contain “architect*” terms. Therefore, we admit that we might have missed certain number of ARPs (i.e., 14.3\%) that do not contain “architect*” terms. We added in our replication package \citep{dataset} the randomly selected posts (i.e., 1,068 posts) and the labeling results (i.e., the 18 ARPs which contain “architect*” terms and the 3 ARPs which do not contain “architect*” terms) for replication purpose.}
 
\textcolor{black}{Another potential threat to the construct validity of this study concerns the manual filtering of ARPs, in which we did not calculate an inter-rater reliability score such as Cohen’s Kappa \citep{cohen1960coefficient}. This may introduce personal bias due to varying interpretations or oversight. As detailed in Section 5.2, the first author independently assessed a random sample of 1,000 ARPs using the criteria in Table~\ref{mainInclusionExclusion}. The second and third authors reviewed the results to ensure a shared understanding. Through this process, the team reached consensus on the filtering guidelines. For 51 controversial cases, we applied a \textit{negotiated agreement} approach~\citep{campbell2013coding} to ensure consistent application of the criteria. 
%To mitigate this risk, we employed a rigorous and collaborative filtering process aimed at reducing the likelihood of including invalid posts. As described in Section~\ref{datacollectionAndFiltering}, during the pilot filtering phase, the first author randomly selected and independently assessed a subset of 1,000 ARPs from the initial set of 13,205 ARPs, based on the inclusion and exclusion criteria defined in Table~\ref{CriteriaForARPsWithRevisionInfo}. The second and third authors subsequently reviewed and examined the assessment results to ensure a shared understanding of the filtering criteria. This collaborative review process enabled the first three authors to reach consensus on the post filtering guidelines. During this process, we identified 51 posts that raised controversies or misunderstandings. These cases were jointly reviewed and resolved through a \textit{negotiated agreement} approach~\citep{campbell2013coding}, ensuring consistent application of the criteria across the dataset. 
We believe the steps taken, including the pilot phase and collaborative resolution of disagreements, have partially mitigated this validity threat.}

\subsection{External validity} 
External validity refers to the extent to which the findings of the study can be generalized in other settings \citep{wohlin2012experimentation}. In this research, we only used SO as the source to investigate the revisions of ARQs. Even though SO is a widely used and popular developer Q\&A site, this unique source still poses a threat to the diversity of the study results. To mitigate this threat, our research could be further enhanced by including more sources (e.g., Reddit). Also, researchers might consider going to the fields and asking for feedback directly from practitioners (e.g., architects and developers) to better understand how they revise ARQs.

\subsection{Reliability} 
Reliability refers to whether the study will provide the same results and findings when it is replicated by other researchers \citep{wohlin2012experimentation}. In this study, the threat to reliability could be related to the process of manual data analysis (e.g., qualitative analysis of ARQ revisions for answering RQ2, RQ3, and RQ4). To mitigate this threat, we (the authors of this study) employed two qualitative techniques (open coding and constant comparison) that are commonly used in empirical studies of software engineering \citep{seaman1999qualitative}. Moreover, as mentioned in Section \ref{dataAnalyis}, the results from data analysis (e.g., categories of missing and further provided information in ARQs) were cross-checked by involving the three authors of the study and any disagreement was resolved by using a negotiated agreement approach \citep{campbell2013coding}. To guarantee the reliability of our results and findings, a replication package, containing the dataset used and the encoded data produced as well as the SQL query used to search ARPs in SO, has been made available \citep{dataset}, allowing other researchers to evaluate the rigor of the design and replicate the study. With these measures, this threat has been partially reduced.
%A substantial threat to the reliability of our results is bias in the labeling process used to generate categories..... To investigate this bias, we conducted a user study with 11 professional developers to check on the appropriateness of categories that are produced in this study. We found that on average across all categories, xxx percent of the classification results are accepted by the majority of the developers, and 73 percent of the classification results are accepted by all developers.

\section{Conclusions}\label{Conclusions}
SO employs various mechanisms (such as the question revision process) to ensure the quality of the information provided in its posts. We conducted an empirical study on SO to explore how SO users revise ARQs. Specifically, we used a qualitative analysis approach to analyze 4,114 ARPs that we filtered from 13,205 ARPs. We intended to identify the prevalence of ARQ revisions, the types of missing and further provided information in ARQs, the purposes of the further provided information in ARQs, and the impact of the further provided information in ARQs on the answers/architecture solutions. The main results are that: (1) a broad range (14 categories) of information is missing when SO users frame or ask ARQs, and \textit{design context} is the most frequently missing and further provided category of information in ARQs after being posted on SO; (2) \textit{clarify the understanding of architecture under design} and \textit{improve the readability of architecture problem} are the two major purposes of the further provided information in ARQs; (3) the further provided information in ARQs has several impacts on the quality of answers/architecture solutions, such as \textit{making architecture solution useful}, \textit{making architecture solution informative}, among others.
 
We believe that our study on revising ARQs can be leveraged in several ways to improve the quality of the information in ARPs. For example, our findings can help researchers and practitioners by knowing what types of information are often missing and further provided in ARQs after they are posted, so that they can develop new techniques and tools to help SO users during the revision of ARQs. In addition, dedicated bots could be provided in SO to warn about ARQs that are unclear or require further clarification. 

In the next step, we plan to \textcolor{black}{(1) investigate the distinguishing features of unrevised ARQs by conducting a comparative analysis between revised and unrevised ARQs in order to identify patterns and attributes associated with question quality (e.g., clarity, completeness)}; (2) examine what types of information are revised and how answers or architectural solutions provided to ARQs are updated on Stack Overflow; and (3) develop a (semi-)automatic approach to detect missing information and assist users in revising architecture-related posts on Stack Overflow.

\section*{Acknowledgments}
This work is partially sponsored by the National Natural Science Foundation of China (NSFC) under Grant No. 62172311 and the financial support from the China Scholarship Council.

\section*{Data Availability Statements}
The replication package of this study has been made available at \citep{dataset}.

\bibliographystyle{spbasic}
\bibliography{references}

\end{document}